\documentclass[12pt]{article}

\usepackage{amsmath,amsfonts,amssymb,latexsym,graphicx,hyphenat}
\usepackage[T1]{fontenc}
\usepackage{textcomp}

\usepackage{caption}
\usepackage{subcaption}
\usepackage{rotating}  

\newtheorem{remark}{Remark}[section]
\newtheorem{example}{Example}[section]
\numberwithin{equation}{section}
\def\be{\begin{equation}}
\def\ee{\end{equation}}
\def\bq{\begin{eqnarray}}
\def\eq{\end{eqnarray}}
\def\beq{\begin{eqnarray}}
\def\eeq{\end{eqnarray}}

\begin{document}
\title{\textsc{Structural stability  and general relativity}}
\author{\Large{\textsc{Spiros Cotsakis}}$^{1,2}$\thanks{\texttt{skot@aegean.gr}}\\
$^{1}$Clare Hall, University of Cambridge, \\
Herschel Road, Cambridge CB3 9AL, United Kingdom\\  \\
$^{2}$Institute of Gravitation and Cosmology,  RUDN University\\
ul. Miklukho-Maklaya 6, Moscow 117198, Russia}
\date{June 2025}

\maketitle
\begin{center}
\begin{minipage}{0.75\textwidth}
    \centering\itshape 
    ``... versality implies the structural stability of the deformation.''

    \hfill --- V. I . Arnold (1994)
  \end{minipage}
  \end{center}
\newpage%

\begin{center}
  \begin{minipage}{0.75\textwidth}
    \centering\itshape 
    ``In order that stationary processes may exist in a real system for a long time, they must be stable not only with respect to the coordinates and the velocities, but also with respect to small variations of the form itself of the differential equations describing the system ... these [small variations of the differential equations] shall be assumed at first to be such as not to vary the order of the initial differential equation ... such as not to force us to reject the idealization connected with the number of degrees of freedom. ... Systems which are such as not to vary in their essential features for a small variation of the form of the differential equations we shall call \emph{``coarse''} systems... .''

    \hfill --- A. A. Andronov, A.A. Vitt, and S. E. Khaikin (1966)
  \end{minipage}

  \vspace{1cm} 
  \begin{minipage}{0.75\textwidth}
    \centering\itshape 
    ``There seems to be a time scale in all natural processes beyond which structural stability and calculability become incompatible ..., and today the physicist sacrifices structural stability for computability.''

    \hfill --- R. Thom (1972)
  \end{minipage}

  \vspace{1cm} 
  \begin{minipage}{0.75\textwidth}
    \centering\itshape 
    ``When does the phase portrait itself persist under perturbations of the vector field? This is the problem of structural stability.''

    \hfill --- M. W. Hirsch and S. Smale (1974)
  \end{minipage}

    \vspace{1cm} 
  \begin{minipage}{0.75\textwidth}
    \centering\itshape 
    ``The word bifurcation ... is used in a wide sense to indicate every qualitative topological metamorphosis of a picture under the variation of parameters on which the object being studied depends.''

    \hfill --- V. I . Arnold (1988)
  \end{minipage}
\end{center}
\newpage
\begin{abstract}
\noindent We review recent developments in structural stability as applied to key topics in general relativity. For a  nonlinear dynamical system arising from the Einstein equations by a symmetry reduction, bifurcation theory fully characterizes the set of all stable perturbations of the system, known as the `versal unfolding'. This construction yields a comprehensive classification of  qualitatively distinct solutions and their metamorphoses into new topological forms, parametrized by the codimension of the bifurcation in each case. We illustrate these ideas through  bifurcations in the simplest Friedmann models, the Oppenheimer-Snyder black hole,  the evolution of causal geodesic congruences in cosmology and black-hole spacetimes,  crease flow on  event horizons, and the Friedmann-Lema\^itre equations.  Finally, we list open problems and briefly discuss  emerging aspects such as partial differential equation stability of versal families, the general relativity landscape, and potential connections between gravitational versal unfoldings and those of the Maxwell, Dirac, and Schr\"{o}dinger equations.
\end{abstract}
\newpage
\tableofcontents
\newpage
\section{Introduction and overview}
\subsection{General relativity and symmetry reduction}
General relativity  occupies a central position in modern physical theories. It naturally interfaces with other areas of theoretical physics, partially `unifying' them via the Einstein equations:   the stress tensor on the  right-hand-side couples directly to  the geometric (gravitational) sector. This coupling yields a rich array of physical predictions and profound effects many of which have been confirmed observationally. Whenever an exact solution of the Einstein equations is available \cite{exact}, one performs a symmetry reduction that converts the Einstein partial differential equation (PDE) into a finite-dimensional system of ordinary differential equations (ODEs). Such reductions --- common in homogeneous cosmology, black-hole problems, relativistic astrophysics, and gravitational-wave studies ---  `collapse' the PDE  to a lower--dimensional dynamical system  (cf. e.g., \cite{ryan}-\cite{cy22}). Although some information is lost, the resulting ODE systems remain of significant  mathematical and physical interest.

\subsection{Analogs in reaction-diffusion and travelling waves}
This symmetry-reduction-related situation is not unique to general relativity, but appears in many important --- and unrelated both physically and mathematically to gravity --- PDEs, notably reaction-diffusion equations, where one derives a `travelling-wave' equation near the wave front. Such travelling wave equations are the exact analogues of the symmetry-reduced equations of general relativity, and while the properties of the reduced PDE cannot capture the essence of the full `unreduced' PDE, most of the reduced equations obtained this way have a common property: they are structurally unstable. This creates an equally amazing variety of phenomena at the reduced-ODE level, and frame the fundamental question of the  ODE-PDE correspondence  (see \cite{1}-\cite{7} for reaction-diffusion and other types of equations).
A main reason for the structural instability of travelling-wave-like equations is their degenerate nature and, as we shall demonstrate below, the same  holds in general relativity. In particular, while the techniques to obtain stability at the PDE level for reaction-diffusion and the Einstein equations are very different, the symmetry-reduced dynamical systems typically obtained by such a process in the case of general relativity are typically  structurally unstable as well. This instability has important implications for gravitational models, yet remains underexplored in the   literature.

\subsection{Structural stability vs.\ instability}
Structural stability  is a well-established mathematical concept concerning the robustness  of dynamical laws under small perturbations of their defining equations, rather than
the stability of particular solutions.  Imposing structural stability  restricts solution behaviour  to only three possible types: stable (perturbations decay), unstable (perturbations grow), or saddle-like (mixed stability). That is, either there is nothing new, i.e., a perturbing solution rapidly dies out and the system returns to its unperturbed state, or diverges, ripping the system apart, or - as perhaps the worst alternative - a mixture of the two that results in an unstable saddle-like state. Consequently, such a system is in some sense always `trapped in itself' being unable to pass to another, qualitatively different, state like for example one of those we observe around us (this will be made precise below). However, structural stability imposes very strong restrictions to the solutions describing different manifestations of the physical phenomena associated with these laws, for if a system is described by such equations then the set of its possible behaviours is restricted by the condition that any nearby system, i.e., any small perturbation of the system itself, behaves exactly as the original one \cite{ant0}-\cite{du0}.

\subsection{Historical background}
While this review builds on the author's recent results \cite{cot23}--\cite{cot25a}, it is rooted in a longer tradition of research spanning four key areas of general relativity:
\begin{itemize}
  \item Linearization (in-)stability of the Einstein equations
  \item Null hypersurfaces, caustics, wavefronts, and event horizons
  \item Singularity theorems and the issue of the focusing state
  \item Dynamical systems in cosmology and relativistic astrophysics
\end{itemize}
Below we briefly survey each area and highlight its connection to gravitational bifurcation theory. A common theme is the evolution problem of the Einstein equations and the physical plausibility of their solutions.

Early work on the \emph{linearization stability} of the Einstein equations laid the foundation for understanding when small perturbations of a known solution again satisfy the full field equations \cite{ycbdeser}. Moncrief's seminal analysis showed that, under suitable symmetry and boundary-condition assumptions, the space of nearby solutions forms a smooth manifold \cite{Moncrief1975,Moncrief1976}. Fischer and Marsden extended this by casting the Cauchy problem into a symmetric-hyperbolic framework and identifying constraint-equation obstructions that block linearized solutions from integrating to exact ones \cite{FischerMarsden1979}-\cite{Fischer1996}. Since linearization stability of the Einstein equations is equivalent to that of their constraint equations \cite{FischerMarsden1979b}, these works pinpointed when one can apply the implicit-function theorem to the constraint map \cite{ycb09}.

The study of \emph{null hypersurfaces, caustics, and wavefronts} dates back to Raychaudhuri's and Sachs's analysis of geodesic-congruence focusing \cite{Raychaudhuri1955}-\cite{Sachs1962}, Penrose's investigation of gravitational lensing and wavefront caustics \cite{Penrose1968}, and Friedrich \& Stewart's Hamiltonian treatment of Legendrian and Lagrangian singularities in Minkowski space \cite{StewartFriedrich1982,FriedrichStewart1983}. These works revealed how singularities in the `optical' equations signal breakdown of smooth null foliations, an early glimpse of dispersive behavior in the characteristic formulation.

The \emph{singularity theorems}  cemented the connection between causal focusing and inevitable spacetime singularities under very general energy and causality conditions \cite{pen65}-\cite{he}. Physically, once a congruence of null or timelike geodesics reaches a caustic, the classical evolution `breaks down', just as a bifurcation in \(\dot x = x^2\) signals a qualitative change in the solution set.

Finally, \emph{dynamical-systems methods} in cosmology and relativistic astrophysics have roots in Misner's mixmaster (Bianchi IX) model and the BKL oscillatory approach to singularities \cite{mis2,bkl2}, later elaborated by Barrow's chaotic-cosmology work \cite{ba82}, and explored by Collins \& Hawking \cite{c71,ch73} and Bogoyavlenski \& Novikov \cite{bogo1}. More recently, the monographs  \cite{bogo,we97,stew1991}, together with Choptuik's discovery of critical phenomena in gravitational collapse \cite{Choptuik1993}, have shown how ODE and PDE bifurcations underlie cosmic evolution and black-hole formation.

Together, these strands sketch general relativity as a theory rich in structural instabilities and pave the way for applying \emph{versal unfoldings} and bifurcation theory techniques to classify every way in which symmetry-reduced gravitational systems can deform and bifurcate, the subject of Sections~\ref{method} and \ref{bifningr}.

\subsection{Bifurcation theory in gravitational context}
Physically speaking, the opposite of structural stability can be said to lie with `bifurcating' systems in which small changes can produce qualitatively new solutions. Bifurcation theory may be defined as the study of structurally unstable, also called `dispersive',  phenomena, and while its precise nature as a mathematical theory is probably unfamiliar to, or perhaps even might be considered as peripheral at best by, experts in the field of gravitation, the purpose of this paper is rather to suggest that there is a fundamental link between gravity and bifurcation theory and explore this relation using methods of the  singularity theory of functions\footnote{We shall use the words structural stability, hyperbolicity, and non-degeneracy as synonymous, and the word bifurcation as synonymous to structural instability, dispersive behaviour, non-hyperbolicity, and degeneracy. We shall also generally use `bifurcation theory' as a generic phrase describing all three mathematical theories of `bifurcations', `singularities' (of functions), and `catastrophes'. For further discussion and proofs of the \emph{general} bifurcation theory statements used below the reader is invited to consult Refs. \cite{ant2}-\cite{ar86}, while for all statements related to \emph{gravitational} bifurcations,  we refer to \cite{cot23}-\cite{cot25a}.}.

There is a great difference between structurally stable and dispersive systems of dynamical equations and for our present purposes a very characteristic manifestation of that difference is provided by the phenomenon of gravitational instability, as for example in the problem of cosmological structure formation where, while it usually treated in the linear approximation as in physical cosmology \cite{weinberg1}-\cite{bau}, the universe was very smooth but has become very inhomogeneous on small scales as cosmological observations clearly suggest. The universe `settles down' to a novel state which albeit  a changing one, it does not suddenly disappear but `persists' for a time comparable to the observed lifetime of the system.

It is a standard geometric procedure in gravitational studies to symmetry-reduce the full Einstein  field equations as applied to specialized situations describing phenomena of interest such as  in cosmology, black holes, and other areas of gravity research. Most commonly, such a reduction is accompanied  by the drastic transformation of the resulting differential systems from partial to ordinary differential equations, in other words reducing an infinite-dimensional degrees of freedom system to a finite-dimensional one, such as is the case for example in a Friedmann-Lema\'{i}tre context,  or a Bianchi-type situation, or similar finitely-dimensional reductions in black hole spacetimes, etc. Such reductions often lead to interesting idealizations from a physical point of view and also to intriguing possibilities of adhering with observations. But while they lead to an obvious partial differential equation stability problem, that is whether the found symmetric `ordinary differential equation' solution is stable or unstable in the full `unreduced' context, they usually also reveal certain important features shared not only by the symmetrically reduced system but also by the original full (i.e., unsymmetric) system of partial differential equations: The symmetrically reduced systems obtained are typically structurally unstable.  This is a common property not only met in the  Einstein equations when applied to concrete symmetric equations, or analogous ones in modified gravity or  string theories,  but also  in `optical' type systems describing the evolution of congruences of causal curves or higher-dimensional gravitating objects in spacetime. The nature of the resulting structural instability of the symmetry-reduced laws can be succinctly summarized both physically and mathematically by the statement that:
\begin{center}
   Gravitation is a bifurcation problem.
\end{center}

Furthermore, gravitational systems share two characteristic features which are in fact very akin to those of typical structurally unstable systems:
\begin{enumerate}
\item They contain parameters whose values are not really known exactly.
\item They are governed by differential equations which are highly sensitive to  small  modifications of their defining terms.
\end{enumerate}
The fluid parameter $\gamma$, or a cosmological constant are examples of the first property, while any `randomly chosen' system of equations describing a gravity problem may be indicative of the second. Although a demand of structural stability is understandable as it implies that a property possessed by such a system is then physically very plausible, such a restriction imposed on an unstable system satisfying the two above conditions may lead to non-characteristic results attributed to the original equations, and leave out other behaviours which might lie closer to the nature of the system's instabilities.

\subsection{Key questions and scope of this review}
Another related issue is what does one really mean by a `stable solution' in the  context of a structurally unstable gravitational system? How may one classify the various unstable modes available in a given gravitational problem described and governed by a structurally unstable set of dynamical equations? Such issues point to a further consideration of related questions. For example,  how can one perturb a structurally unstable gravitational system? What is the possible range of such perturbations, and how are the solutions of the original system  being affected by the latter? What is the meaning of the phrase  `small perturbation' for a structurally unstable situation where there may be almost ignorable perturbations that have the ability to completely disrupt all orbits? Most importantly, how can one `lift' all possible solution branches appearing at the level of the symmetry-reduced finitely-dimensional dynamical system to the full un-symmetric equation (say, the Einstein equations) and decide which one is finally picked?

The main message of this paper is that gravitation is a  structurally unstable phenomenon and must be treated using an approach where this feature does not disappear altogether in the process but persists. In particular, the observed cosmic structures could have never been formed and persisted had gravity been structurally stable. How does then gravitational instability in its various forms manifest itself and related to the structural instability of the governing gravitational equations? How can one describe the latter in a mathematically precise yet physically consistent and plausible way? What are the main physical implications predicted by gravitational structural instability alone that could never have arisen had we treated the equations of a particular gravitational setup as being structurally stable?

We shall return to some of these questions in the coming sections of this work, where we shall introduce the approach of bifurcation theory in the general framework of gravitational physics as a possible solution to this problem. Basically, very roughly speaking, this approach to the problem of general gravitational instabilities dictates that: one must first clearly recognise the kind or degree of  degeneracy of the original system by bringing it to a `normal form'; secondly,   embed the system of dynamical equations in a suitable larger parametric family customized to the given problem; and thirdly, prove that this family contains generically all possible stable perturbations of the original system. This is a long path, but eventually becomes very rewarding.  One acquires control over all degeneracies of the original system, now in a structurally stable framework where each unstable mode persists, and in this way one is led to the complete spectrum of all possible behaviours associated with the original system.

\subsection{Outline of the paper}

Section \ref{method} gives a brief exposition of the main methods of bifurcation theory. In Section \ref{hyp-disp}, we discuss the main differences between hyperbolic and nonhyperbolic systems, and in the remaining of Section \ref{method} we detail on the main mathematical tools used in such an approach, with an eye to their subsequent application in a gravitational context (cf. Section 3). In Section \ref{bifningr}, we  study the picture which emerges from an application of these methods to some of the most basic and well-studied gravity problems in general relativity, cosmology, and black holes. In particular, in Section \ref{central} we formulate the main problem addressed in this work and write down the central objectives in the search of gravitational bifurcations, while the remaining of this Section presents the main features of an application of bifurcation theory to certain important problems formulated after symmetry reduction of the Einstein equations. We also present the main results of such an analysis as applied  to the Raychaudhuri equation and related systems, and Section \ref{open}  comprises a representative list of open problems associated with this approach. In the last Section, we discuss more general issues which emerge from gravitational bifurcations.

While this work may be viewed as a review based on the results of Refs. \cite{cot23}-\cite{cot25a}, however, we present here  a more qualitative way towards those results, with a view to the bigger picture which necessarily emerges and relates to a forming interface between general relativity and bifurcation theory. We consequently skip all mathematical details of the long arguments and calculations needed to prove many of the statements made in those references,  but only referred to here in a `high-level' manner. After reading the present paper, we hope that the reader will be sufficiently motivated to at least look at some of the proofs in those references.

\section{Dynamics and bifurcations}\label{method}
In this Section, we   describe very briefly some of the main methods in the vast territory of bifurcation theory in an effort to orient the reader in the maze of details and also to establish the notation  we shall use subsequently in Section \ref{bifningr}. We have developed  the introductory material in the  first few subsections,  while later in this Section we shall discuss additional, somewhat more advanced, methodology topics.

Consider a set of dynamical equations that govern the evolution of a gravitating system, we shall call this set \emph{the original problem}. For example, we imagine that this problem can be described mathematically by an autonomous dynamical  system obtained from the Einstein equations after some symmetry idealization has been imposed, like with some of the problems analyzed in Section \ref{bifningr}.
\subsection{Hyperbolic vs. dispersive systems}\label{hyp-disp}
\subsubsection{General comments}
Besides the theory of small fluctuations mentioned in Section 1 in the context of physical cosmology, an essentially hyperbolic approach, i.e., assuming structural stability in the context of individual systems, is also used in other, somewhat more mathematical,  treatments of gravity problems, cf. e.g., the pioneering works \cite{ryan}-\cite{cop18} and refs. therein, and has its roots in the standard theory of (hyperbolic) dynamical systems \cite{ant0}-\cite{du0}. While this approach imposes two very important conditions on the gravity systems studied in this way,  nevertheless it is the standard, well-known path that leads  to systems with (usually complicated) hyperbolic  equilibria: firstly, near such equilibria the equations become topologically equivalent to their linear parts, and secondly, `nearby' nonlinear systems behave like their unperturbed counterparts.

In this work, we shall instead consider structurally unstable systems only, and  follow their evolution as the parameters present in the problem vary. Our approach has a very high degree of applicability not only in cosmology but in other areas as well (for a description of possible cosmological areas of application of these ideas, cf. e.g., \cite{cy22}), because - as we hope it will become clear in the following - the abundance of such systems in gravitation is  high to such an extent as it may not be an exaggeration to say that a gravity problem obtained by a symmetry reduction of the Einstein (or `similar') equations to a dynamical systems form is \emph{typically} structurally unstable.

To motivate  gravitational bifurcation theory employed in this work, let us recall the ways in which difference between a hyperbolic and a non-hyperbolic system arises (a hyperbolic system is one whose all equilibria, i.e., steady state solutions, are hyperbolic).  Equilibria of a hyperbolic system are always isolated and the behaviour of solutions near them is qualitatively the same as that of their linear parts. However, dispersive systems may exhibit \emph{bifurcating} behaviour and as a rule comprise multiple solutions intertwined with dramatic changes in the behaviour of the orbits. In addition, their global dynamics  may be fully described by `restricting' to only those dimensions which correspond to the non-hyperbolic eigenvalues (i.e., tangent to the null eigenspaces).
\subsubsection{Hyperbolic dynamics}
Let us now explain this difference in a little more detail. Let $x(t)$ be a vector function of the time and consider a gravitating system described by an autonomous dynamical system in finite dimensions,
\be\label{f-sys}
\dot{x}=f(x,\lambda),
\ee
which also depends on some parameters collectively denoted by $\lambda$, as will be the case with most of the gravity problems to be analysed below.
In a hyperbolic approach, the system is arranged so that, for some range of $\lambda$, it has a hyperbolic equilibrium at $x_0$, i.e., none of the eigenvalues of the jacobian of its linear part $D_x f(x_0,\lambda)$ lie on the imaginary axis (including the origin)\footnote{in reality the system may have a number of such equilibria in which case one applies the same method to each one of them.}. Here and throughout we view the equilibrium as a function of a parameter~$\lambda$.  As $\lambda$ varies, as we shall see below,  any eigenvalue whose real part passes through zero marks a critical value $\lambda_c$ at which the equilibrium ceases to be hyperbolic (i.e.,  becomes `marginally' or `neutrally' stable in other terminology) and typically undergoes a bifurcation. An equilibrium \(x_0(\lambda)\) is \emph{hyperbolic} if no eigenvalue \(\mu(\lambda)\) of \(Df(x_0(\lambda))\) lies on the imaginary axis.  Equivalently, as \(\lambda\) approaches a critical value \(\lambda_c\) at which  the real part \(\Re\,\mu(\lambda_c)=0\), the equilibrium becomes neutrally stable, signalling a loss of hyperbolicity and typically the onset of a bifurcation.

A hyperbolic equilibrium solution \emph{persists}, as does the nature of its stability for solutions $x$ near $x_0$ when moving to nearby systems, which as a result share one of the three possible behaviours in this case, namely,
\begin{itemize}
\item nodal,
\item focus-like, and
\item saddle-like.
\end{itemize}
The first two can be either stable or unstable, while (linear and nonlinear) saddles are unstable and contain  a mixture of stable or unstable orbits. More succinctly, the possible behaviours in hyperbolic dynamics are:  saddles, sinks, or sources.

This simplicity in the resulting behaviour of  hyperbolic systems appears because in this case the Jacobian matrix $D_x f(x_0,\lambda_0)$ is invertible for any $\lambda_0$ in the said range. Therefore the implicit function theorem guarantees the existence of an open interval \(I\ni\lambda_0\) and a unique smooth function
\[
  x\colon I\to\mathbb{R}^n,
  \quad x(\lambda_0)=x_0,
\]
such that
\[
  f\bigl(x(\lambda),\lambda\bigr)=0
  \quad\text{for all }\lambda\in I.
\]
So  by continuity of the eigenvalues of $D_x f(x,\lambda)$ with respect to $\lambda$ sufficiently close to $\lambda_0$, the matrix $D_x f(x,\lambda)$ also has no eigenvalues on the imaginary axis. As we shall see below, this simplicity does not carry over to the non-hyperbolic case.

\subsubsection{Dispersive dynamics}
Now suppose one can show that the system (\ref{f-sys}) is structurally unstable (`nonhyperbolic', or `dispersive') and has a nonhyperbolic equilibrium at $(x_0,\lambda_0)$.
Then the non-invertibility of the Jacobian  implies a \emph{necessary} condition for the existence of multiple solutions $x(\lambda)$ of the equation $f(x,\lambda)=0$ (instead of a unique solution as before). To see this, suppose instead that
\[
  \det D_x f(x_0,\lambda_0)=0,
\]
so that the Jacobian matrix $D_x f(x_0,\lambda_0)$ is singular.  Then $(x_0,\lambda_0)$ is a \emph{nonhyperbolic} (or \emph{dispersive}) equilibrium and the implicit function theorem no longer guarantees a unique branch.  In particular, the failure of invertibility at $\lambda_0$ is a \emph{necessary} condition for the coexistence of two or more smooth solutions $x(\lambda)$ to the equation  $f\bigl(x(\lambda),\lambda\bigr)=0$  near $\lambda_0$, and thus signals a potential local bifurcation.

To provide a solid basis for future intuitive interpretations, we may introduce the `landscape' as the graph of all equilibrium solutions
\[
\{\,(x,\lambda)\mid f(x,\lambda)=0\},
\]
in the combined $(x,\lambda)$-space. This set may represent the equilibrium set for a given dynamical system (cf. Section 2.2). For most $\lambda$ this graph looks like a single smooth surface (a unique branch~$x(\lambda)$).
In this landscape analogy, a non-hyperbolic equilibrium corresponds to a valley floor that has been `flattened' (i.e.\ the Jacobian becomes singular), and suddenly one \emph{sheet} of the surface can split into two or more sheets, each sheet is a new equilibrium branch.  Instead of a single well with a unique bottom, the flattened region harbors several adjacent basins.  A small perturbation can then send the system rolling into any one of these dips -- each representing a distinct smooth solution branch \(x(\lambda)\).  Thus, the singular Jacobian signals the emergence of multiple equilibria.

Hence, the intriguing possibility arises that there are radically new, essentially nonlinear effects emerging for the system (\ref{f-sys}) near $(x_0,\lambda_0)$.
This means that new solutions may be created or destroyed, and periodic or quasi-periodic behaviours may arise.
We shall call a point $(x_0,\lambda_0)$ with,
\be \label{singular}
f(x_0,\lambda_0)=0,\quad D_x f(x_0,\lambda_0)=0,
\ee
a \emph{singularity}\footnote{No relation to `spacetime  singularities'. We hope that the probably somewhat excessive use of this term usually associated with the big bang and black holes  does not preclude its usage also in the  present context. It is interesting that in the present context spacetime singularities do not typically arise.}. This definition implies that a singularity is just a nonhyperbolic point. We note, however,  that a singularity need not be a \emph{bifurcation point}: If we let $n(\lambda)$ denote the number of solutions $x$  for which $(x,\lambda)$ is a solution of the equation $f(x,\lambda)=0$, we say that $(x_0,\lambda_0)$ is a \emph{bifurcation point} of the system (\ref{f-sys}) if it is a singularity and also $n(\lambda)$ changes as $\lambda$ varies in a neighborhood of $\lambda_0$. Intuitively, a bifurcation point is where the landscape not only flattens but actually reshapes: as \(\lambda\) passes through \(\lambda_0\), one valley floor can split into two (or two merge into one), so that the number of available basins -- i.e.\ the number of equilibrium branches -- changes abruptly.

Most importantly, in this case the phase portraits of the system upon parameter variation undergo continuous transformations, `metamorphoses' to new forms accompanied by topologically inequivalent types of behaviour. An important lesson to be learnt from this is that for a structurally unstable system of dynamical equations, nearby vector fields can have very different orbit structure. This means that when at a certain parameter value $\lambda_0$ the vector field $f(x,\lambda_0)$ has a nonhyperbolic equilibrium where new solutions are then created as the parameter $\lambda$ is varied, one should not only study the orbit structure of the system $f(x,\lambda_0)$ near $(x_0,\lambda_0)$ as is  done in a hyperbolic approach to studying gravity problems, but also \emph{the local orbit structure of nearby systems}, that is the systems $f(x,\lambda)$ for $\lambda$ near $\lambda_0$ (this is a very important remark usually ignored, however, cf. \cite{ar83}, p. 224).

For example, consider the simplest case of a bifurcating system where one shows that $D_x f(x_0,\lambda_0)$ has a \emph{single} zero eigenvalue and the remaining ones having nonzero real parts. Intuitively, if we think of eigenvalues as vibration frequencies of the landscape, then a zero eigenvalue is a `zero-frequency' mode, i.e.\ no oscillation at all. In our terrain picture it corresponds to a perfectly flat direction: no slope to pull the system one way or the other, and no restoring force to push it back. Anything that moves a bit along that direction simply stays where one puts it, reflecting the neutral (centre manifold) dynamics along that flat ridge. (To push this interpretation a bit further (when crossing the imaginary axis), a conjugate pair of purely imaginary eigenvalues indicates neutral, oscillatory dynamics, a \emph{center} in phase space.  In our landscape analogy, this corresponds to a perfectly circular, frictionless bowl: a marble given a small nudge will perpetually orbit around the rim at constant amplitude, neither spiraling in (decaying) nor flying out (growing).  In other words, the landscape curvature in orthogonal directions provides just the right `centripetal' pull to sustain continuous, undamped oscillations.)

What is then the nature of the equilibrium solution $(x_0,\lambda_0)$? In this connection, a  notable feature of structurally unstable systems is that the laws governing them may be recast in a `pure' form devoid of any hyperbolic behaviour. In this case, the number of equations is drastically reduced to only the essential ones, and their dynamics is suitably restricted  to be on a phase space of \emph{smaller} dimensionality, the so-called `centre manifold' (unlike in the structurally stable, hyperbolic class where this reduction never occurs).
In this simplest case, the centre manifold is 1-dimensional, and the problem for a $k$-dimensional family ($k$ being the components of the parameter $\lambda$) is drastically reduced to evolve on a 1-dimensional centre manifold. Intuitively, a centre manifold is a `flat plateau' through a non-hyperbolic equilibrium -- the collection of directions in which neither strong restoring nor repelling forces act, so trajectories linger there, neither growing nor decaying, so motion is slow and `marginal'.  By restricting the dynamics to this lower-dimensional plateau, one thus captures the slow, critical modes that actually govern how the system will split into new branches as parameters pass through a bifurcation point.

All arguments below include the fact  that a centre-manifold or `Lyapunov-Schmidt' reduction (to which we shall return later for more details) on some manifold of small dimension (i.e., the `centre manifold') is  performed for the original problem in question, as a single step in the whole process (before the purely `bifurcational' parts  in the procedure take place). In other words, in the case of dispersive systems the reduced dynamics on the centre manifold is along the manifold tangent to the eigenspace corresponding to the dispersive eigenvalues, a vast simplification. In all gravitational examples studied in this work, the centre manifold turns out to have dimension up to two.

\subsection{Dynamical systems formulation}\label{A}
We assume that there is a change of variables that brings the differential equations that define the original problem in the following \emph{dynamical system form}\footnote{We shall assume below that this form of the equations always corresponds to an \emph{autonomous} dynamical system in a finite number of dimensions. Specifically, while non-autonomous systems sometimes appear in cosmology and gravity, and can be reduced to a corresponding autonomous system by raising the dimensionality by one, analogous but different techniques are required for their efficient treatment, and these lie outside the scope of this review. Also we shall not consider the important subject of \emph{maps} (i.e., discrete dynamical systems), which although  can be treated with similar techniques,  their main applications to cosmology (e.g., chaotic behaviour) require a separate extensive treatment using the present methods.},
\be \label{w-sys}
\dot{w}=G(w,\lambda),
\ee
where $G$ is a suitably smooth vector field of class $C^r$ (in most cases $r\geq 4$ will do). Here $w\in U\subset \mathbb{R}^n$,  and $\lambda\in B\subset \mathbb{R}^k$  is a $k$-vector of parameters, we shall call the domain $U$ \emph{phase space}, $B$ the \emph{space of parameters}, and $G$ a \emph{family} of vector fields on $U$ with \emph{base} $B$. (We shall avoid technical jargon as much as this is feasible without being led to ambiguities, bearing in mind that this subject very quickly gets very technical.)

The first step is then to find all (non-hyperbolic) equilibria of Eq. (\ref{w-sys}). This, regardless of often stated in a gravitational setting as an easy exercise in certain cases, is by its very nature a problem in algebraic geometry, defined as being the study of systems of algebraic equations of several variables like the right-hand-side of Eq. (\ref{w-sys}) equal to zero, $G(w,\lambda)=0$, and the structure one can give to their solutions.

Another reason we add this as a separate step belonging to the bifurcation sequence detailed in this Section is to emphasize the fact that for our purposes it is important that the new variables $w$ that bring the original system of equations to the form (\ref{w-sys}) are functions of only the `phase space' variables of the original problem, say, $y$, i.e., $w=w(y)$, and \emph{not} of the distinguished parameters of the problem (when they exist). In other words, we need to ensure that any parameter present in the original problem survives in the form (\ref{w-sys}) and is not absorbed in the transformation.

For example, in a Friedmann cosmology context, a variable like $\Omega_\Lambda=\Lambda/3H^2$,  while important in certain hyperbolic analyses (cf. e.g., the interesting analysis in Ref. \cite{pu09}, p. 146 and refs. therein), is not suitable for our present purposes, because while $\Lambda$ was a distinguished parameter in the original Friedmann equations and thus part of the base $B$ of this problem, is now absorbed in the new phase space $U$-variables.

\subsection{Jordan form of the linear part}\label{B}
Let us suppose that (\ref{w-sys}) has a nonhyperbolic equilibrium at $(w,\lambda)=(w_0,\lambda_0)$ (in case of more than one nonhyperbolic equilibria, we should examine each one of them separately).  We shall assume for simplicity that there are no distinguished parameters (the other case is somewhat more complicated, but the main ideas are qualitatively similar).

Moving the equilibrium to the origin with a linear change to new variables $v=w-w_0$ and splitting off the linear part of the resulting vector field, we find
\be
\dot{v} =DH(0)v+\bar{H}(v),
\ee
where we have set $\bar{H} (v)=H(v)-DH(0)v$, and $H(v)=G(v+w_0)$, while $DH(0)$ represents the Jacobian matrix of $H$ evaluated at the origin.

Then passing to further variables $X$, with $v=TX$, where $T$ is the matrix that puts $DH(0)$ into real Jordan canonical form, we finally find the following system,
\be \label{X-sys}
\dot{X}=JX+F(X),
\ee
where $J$ is the Jordan normal form of $DH(0)$, and $F(X)=T^{-1}\bar{H}(TX)$ is of order O(2), that is of second order in the field variable ($X$ in this case). We note that here $X\in\mathbb{R}^n$, and in the system (\ref{X-sys}) we have simplified the linear part of the vector field as much as possible.

\begin{remark} In the case of the presence of a $k$-parameter $\lambda$ entering linearly in the equations, as is commonly the case for instance in some problems with a cosmological constant, there will be an extra linear term in Eq. (\ref{X-sys}) of the form $\bar{K}\lambda$, where $\bar{K}=T^{-1} K$, and $K= D_\lambda H(0,0)$.
\end{remark}

Once we have completed the steps in the previous two subsections and arrived in the form of the original problem given by Eq. (\ref{X-sys}), we have already gained new and valuable information about the original problem. This is so because we may now view it as \emph{a point in the space of all dynamical systems}, where the `singular cases' belong to the boundary hypersurfaces of the domains of the generic cases (i.e., of the structurally stable ones). The singular hypersurfaces have their own singularities, often appearing as intersections of two or more such boundary manifolds. While one may perturb by an arbitrarily small perturbation an individual degenerate system off the singular hypersurface to which it belongs to become a generic (i.e., structurally stable) system without degeneracies, it is impossible to make all degeneracies of a \emph{family} of degenerate systems simultaneously disappear.
Such families appear in gravitational problems as a result of treating their degeneracies, and the number of different parameters required to fully describe their degeneracies is a measure of the degree of complexity, called the \emph{codimension} of the degeneracies of the system in question. The higher the codimension, the greater the complexity of the original problem.
\begin{example}
Consider two different nonlinear gravity problems for which the application of the previous two subsections gives the following two results for the Jordan forms $J_1,J_2$ of their linear parts respectively when we write them in the form of Eq. (\ref{X-sys}):
\be
J_1=\left(
  \begin{array}{cc}
    0 & 0 \\
    0 & -1 \\
  \end{array}
\right),\quad J_2=
\left(
  \begin{array}{cc}
    0 & 1 \\
    0 & 0 \\
  \end{array}
\right)\label{3}.
\ee
We note that both $J_1,J_2$ are non-hyperbolic. Which of the two gravity problems with linear part as above is more degenerate? The degree of degeneracy is roughly given by the number of zero eigenvalues in the Jordan forms, the higher this number the greater the degeneracy, and in this case, it is one for the linear part $J_1$ and two for $J_2$. In the first (resp. second) case, we require one (resp. two) parameter to make up the family (so as to reach any linear system in a small neighborhood of the non-hyperbolic linear system), so the codimension is one in the first and two in the second case.
\end{example}

Being degenerate also means that a system can be further simplified. For instance, one can lower the dimension of the phase space by `neglecting' inessential i.e., hyperbolic variables present in the system (\ref{X-sys}). This is done in the next step.

\subsection{Centre manifold reduction}\label{C}
We first note that for $X=(x,y,z)$, the Jordan matrix $J$ in Eq. (\ref{X-sys}) is a block matrix $J=(A,B,C)$, where $A$ is a $(c\times c)$-matrix having eigenvalues with zero real parts in the `centre' $x$-dimensions, $B$ is a $(s\times s)$-matrix having eigenvalues with negative real parts in the `stable' $y$-dimensions, and $C$ is a $(u\times u)$-matrix having eigenvalues with positive real parts in the `unstable' $z$-dimensions, counting algebraic multiplicity (where $n=c+s+u$). (This is proved in great detail in Refs. \cite{hs,arny78}.) Let us suppose for simplicity that there are no unstable dimensions $z$.

Then the \emph{centre manifold reduction theorem} (cf. e.g., \cite{carr}, chaps. 1, 2) states that:
\begin{enumerate}
  \item near the (nonhyperbolic) origin there exists an invariant graph (manifold) $y=h(x)$, the centre manifold, such that the dynamics given by Eq. (\ref{X-sys}) is equivalent to the dynamics of the reduced $c$-dimensional system,
\be \label{4}
\dot{x} =Ax+f(x,h(x)),
\ee
where $f$ is the $x$-component of the nonlinear part $F(X)$ in Eq. (\ref{X-sys}), that is depending only on the dispersive dimensions $x$;
  \item the two solutions, namely, $x(t)$ on the centre manifold and $(x(t),y(t))$ of the system (\ref{X-sys}) are identical up to exponentially small terms, while their stabilities are the same; and
  \item  the centre manifold can be computed to any finite order of $x$ for small solutions by using the `tangency condition': $\dot{y}-Dh(x)\dot{x}=0$.
\end{enumerate}
When the unstable dimension $z$ is included the reduction result remains the same (except of the stability part of it which is now saddle-like). When a $k$-parameter $\lambda$ is present, the previous discussion persists around the nonhyperbolic origin $(0,0)$, but we obtain an augmented $(c+k)$-system in the form,
\be\label{5}
\dot{x} =Ax+f(x,h(x,\lambda),\lambda),\quad\dot{\lambda}=0.
\ee

The centre manifold reduction theorem implies a reduction of the dynamics of Eq. (\ref{X-sys}) down to acquiring the dimensionality of the centre manifold ($c<n$). (We note that the reduction theorem with $\lambda$ goes through similarly because the added $k$-dimensions are along the centre directions and have no dynamics.)
For example, suppose we have a gravitational system which is brought to the form (\ref{X-sys}) after the application of subsections \ref{A}, \ref{B}, with the $n$-dimensional Jordan matrix $J$ being of the form of the matrix (\ref{3}a) having one zero and $(n-1)$ eigenvalues equal to $-1$. Then the reduction theorem implies that this problem can be reduced to a \emph{one-dimensional} system of the form Eqns. (\ref{4}) or (\ref{5}) depending on the existence of distinguished parameters in the original system.

Bifurcation theory assumes that all systems are already  reduced to a $c$-dimensional centre manifold by the method of this subsection, and deals with dynamical equations in the form (\ref{X-sys}), or, that in Eq.  (\ref{5}) (that is after centre manifold reduction). For various centre manifold calculations in general relativity cf. \cite{cot23}-\cite{cot25a}.

\subsection{Normal form}
In subsection \ref{B}, we simplified the linear part of the vector field and now we proceed to simplify the nonlinear part $F(X)$ in Eq. (\ref{X-sys}). Let us assume the matrix $J$ has $n$ distinct eigenvalues $\nu_1,\dots,\nu_n$ (more general/degenerate cases, like for instance those corresponding to Eqns. (\ref{4}), (\ref{5}), are treated in a more elaborate but essentially similar process). We may think of each eigenvalue $\nu_i$ in the vector as setting the `pitch' or curvature of the local landscape in one principal direction.

We say that the eigenvalue vector $\nu=(\nu_1,\dots,\nu_n)$ is \emph{resonant} if there is a relation of the form $\nu_s=m\cdot\nu$, where $m=(m_1,\dots,m_n)$, $m_i\geq 0,\sum m_i \geq 2$, with all $m_i$ integers. The number  $\sum m_i$  is called the \emph{order} of the resonance, otherwise $\nu=(\nu_1,\dots,\nu_n)$ are \emph{nonresonant} (e.g., $\nu_1=2\nu_2$  is a resonance of order 2, but $2\nu_1=3\nu_2$ is not a resonance). Intuitively, a nonresonant vector means no simple integer-ratio relations exist among those pitches, and each direction vibrates at its own distinct frequency.  In the landscape picture, we can straighten out (normal-form) each slope independently, leaving a neat, decoupled bowl in which nonlinear wrinkles can all be ironed flat. On the other hand, a resonant eigenvalue vector means two or more pitches stand in a simple ratio (e.g.\ one pitch is twice another, like in the example above), so tilting or curving the terrain along one axis inevitably tugs on another.  Geometrically, this creates persistent ridges or troughs that cannot be removed by any smooth change of coordinates, and therefore certain nonlinear couplings remain in the normal form.

The normal form method of simplification of nonlinear vector fields after centre manifold reduction is essentially due to H. Poincar\'e who observed that if in Eq. (\ref{X-sys}) the eigenvalues of $J$ are nonresonant, then (\ref{X-sys}) is equivalent to its linear part after a smooth change of the unknown $X$. This is what happens in the hyperbolic case where we have the best possible conclusion for topological equivalence, i.e., `Hartman-Grobman', and was the fundamental result in Poincar\'e's Thesis.
When, however,  the eigenvalues of $J$ are resonant, one Taylor expands $F(x)$, and by a sequence of analytic transformations of the form $X=Y+h_r (Y),r=2,3,\dots,$ Eq. (\ref{X-sys}) becomes,
\be\label{6}
\dot{Y} =JY+F_2^{\textrm{res}} (Y)+F_3^\textrm{res} (Y)+\dots+F_{r-1}^\textrm{res} (Y)+O(r),
\ee
where one has succeeded in annihilating all nonresonant terms at each order up to order $r$ (this is the meaning of the word `simplification'). This is accomplished by solving a certain linear equation for the $h_r (Y)$ (the `homological equation'), and at each consecutive order starting with the second, the terms $F_i^\textrm{res} (Y),i=2,3,\dots$, referred to as \emph{resonance terms}, represent the remaining \emph{unremovable} terms after the smooth transformation to the $Y$-coordinates at the given order.
Ideally one wants to remove all inessential (i.e., nonresonant) terms at a given order, and proceed to the next, the final result in Eq. (\ref{6}) contains only the essential terms.

The structure of these remaining terms is entirely determined by the linear part of the vector field, while simplifying the terms at order $r$ does not modify any lower-order terms.
One thus obtains an equivalent bifurcation problem to the original, a `normal form' of the field at some desirable order, by constructing at that order a basis of vector fields in the space of homogeneous polynomials which cannot be eliminated. Any term present in the vector field in Eq. (\ref{X-sys}) (or (\ref{4})) that is not among these unremovable terms of the normal form so constructed can be eliminated.

Intuitively, normal forms are like replacing the intricate topography near a flattened valley floor with a simple, canonical shape, such as a parabola for a saddle-node, a symmetric fork for a pitchfork, or a cusp, for capturing the essential way the landscape bifurcates.  By bringing the system into one of these standard templates via a smooth change of coordinates, one immediately sees how many wells will appear, merge, or split as parameters cross the critical value, without being distracted by higher-order wrinkles in the original terrain. The ease of reducing a landscape to a simple normal form hinges on whether its principal curvatures (i.e.\ `eigenvalue pitches') resonate.  In the nonresonant case, each slope can be flattened independently, so the terrain near a bifurcation collapses to a neat, decoupled bowl or fork, \emph{and the normal form contains only the simplest nonlinear terms}.  By contrast, when pitches resonate (e.g.\ one direction's curvature is an integer multiple of another's), any attempt to smooth one slope inevitably distorts its partner, leaving `stubborn' ridges or troughs in the landscape.  Normal form theory then \emph{preserves exactly those irreducible coupling terms}, yielding a canonical shape that faithfully represents the persistent interplay between resonant directions.

This ingenious process results in a new vector field which is in many ways much simpler than that given originally and has important applications (most famously in the Kolmogorov-Arnold-Moser theory of quasiperiodicity). For instance, \emph{any} 2-dimensional vector field that has the matrix $J$ as in (\ref{3}b) as its linear part, has normal form $(0,x^2\pm xy)^T+O(3)$. (This normal form was first discovered by Takens in 1974 and Bogdanov in 1975.)
\begin{example}
For the Oppenheimer-Snyder vector field, $J(x,y)^T+(0,-3/4 y^2 )^T$, one can show that the second-order term is nonresonant and so by the above process can be eliminated, cf. \cite{cot24a}. Therefore there are no second-order terms in the normal form, and one needs to proceed with a computation of the 3rd-order terms. Setting $X=(x,y))$, the OS vector field becomes,
\be
\dot{X}=JX+O(3),
\ee
and the question arises by what process one can determine the nature of the higher-order terms which necessarily arise.
\end{example}
\subsection{Versal unfoldings}
Although as we showed a centre-manifold-reduced system can be further simplified, the normal form obtained this way will still be (structurally) unstable with respect to different perturbations, that is with respect to nearby systems (i.e., vector fields), and so its flow (or, equivalently, that of the original system) will not be fully determined this way.
It is here that bifurcation theory makes its decisive entry, in that it uses the normal form to construct a new system, the \emph{(uni-)versal unfolding} (or, \emph{deformation} in other terminology), that is partly based on the normal form but also contains the right number of parameters needed to take into full account \emph{all} possible perturbations and degeneracies of the normal form system (and so also of the original one).

Intuitively, we may think of a versal unfolding as the master control panel for a singular landscape, a flattened valley floor equipped with a minimal set of `control knobs' (the `unfolding' parameters) that let you tilt, twist, and stretch the terrain in every direction. By adjusting these knobs you reproduce any small, structurally stable (i.e.\ hyperbolic) deformation of the original system, generating all qualitatively distinct ways the valley can split into wells, saddles, or ridges. Thus, no stable perturbation lies outside this universal, parameterized family. In versal unfolding theory, the number of control parameters -- the knobs on our master panel -- equals the \emph{codimension} of the singularity.  Geometrically, codimension is the number of independent degeneracy conditions you must lift to restore hyperbolicity.  In the landscape picture:
\begin{itemize}
\item A codimension-1 singularity (one zero eigenvalue) is a single flattened direction, so one knob suffices to tilt the floor into a generic well or saddle (e.g.\ a saddle-node).
\item  A codimension-2 singularity (two independent zero modes) has a two-dimensional flat plateau, requiring two knobs to span all small deformations.
\item  More generally, codimension \(k\) means \(k\) independent `flat' directions, and thus exactly \(k\) parameters in the versal unfolding to capture every structurally stable nearby terrain.
\end{itemize}
Armed with this picture of intuitive interpretation for the notions of unfolding and codimension, we may now proceed to a more rigorous deployment of these crucial ideas.

Consider an equation of the form $g(x,\lambda)=0$, where we assume that  $x\in \mathbb{R}^n,\lambda\in\mathbb{R}^l$, for given positive integers $n,l$, and we  imagine it represents the equilibrium solutions of a system like  Eq. (\ref{X-sys}) (or, perhaps better (\ref{5})), i.e., we focus on the right-hand sides of them equal to zero, \emph{after} the system has been put into its normal form.  A $k$-\emph{parameter unfolding} of (the normal form) $g$ (of the original problem) is a smooth (usually infinitely-differentiable) family $G(x,\lambda,\mu)$, for the $k$-parameter $\mu$ such that $G(x,\lambda,0)=g(x,\lambda)$. This is a perturbation of $g(x,\lambda)$ since,
\be
G(x,\lambda,\mu)=g(x,\lambda)+(G(x,\lambda,\mu)-G(x,\lambda,0)).
\ee
We say that $G(x,\lambda,\mu)$ is a \emph{versal} unfolding of $g$ if the following crucial property holds: For any small perturbation $p(x,\lambda)$ of the equation $g(x,\lambda)=0$,  that is equations of the form  $g(x,\lambda)+p(x,\lambda)=0$, there is a value of the parameter $\mu_*$ such that,
\be \label{7}
G(x,\lambda,\mu_*)=g(x,\lambda)+p(x,\lambda),
\ee
in the sense of qualitative equivalence. That is, if a versal unfolding $G(x,\lambda,\mu)$ exists, then any small perturbation of the equation $g(x,\lambda)=0$ is \emph{contained} in the the extended function $G(x,\lambda,\mu)$. Therefore this provides a global way to control all small fluctuations of the original problem (which has already been put in normal form). We thus arrive at the \emph{versal unfolding equation} corresponding to the original problem, that is the parametric dynamical system,
\be\label{8}
\dot{x} =G(x,\lambda,\mu).
\ee
Setting $\mu=0$ in Eq. (\ref{8}) we are back to Eq. (\ref{X-sys}) (or, (\ref{5})), that is to the original system written in the suitable dynamical system form that we motivated in Section \ref{A}. We call the $\mu$ an \emph{auxiliary}, or \emph{unfolding} (vector) parameter to differentiate from the \emph{distinguished} vector parameter $\lambda$ possibly present in the original system. The number $k$ of components of the unfolding parameter $\mu$ is called the \emph{codimension} of $g$.
\begin{remark}
We note the following important observation with respect to the versal unfolding construction. While the versal unfolding equation (\ref{8}) deploys smooth relations (e.g., between the auxiliary parameters $\mu$ and the perturbations $p$), there is a great deal of \emph{non-smooth} behaviour (in other terminology `rough' behaviour) in the study of the versal unfolding dynamical system (\ref{8}). This follows because  it is not possible to express $x$ as a smooth solution of $\lambda,\mu$ in the equation $G(x,\lambda,\mu)=0$.
\end{remark}
We shall work extensively  with versal unfoldings chiefly because they are so natural: they are forced to us not as a kind of modification or  speculation expected to be connected to a projected need for a physical extension associated with the original problem, but simply because they must be here because  the  inner structural properties of original system so dictate.  While proving the \emph{versality} of an unfolding is perhaps the most crucial aspect of bifurcation theory, versal unfoldings in addition to the above property of  controlling all small perturbations share a number of other benefits, including the determination of novel global (or quasi-global) properties using only local methods, as well as providing a fresh way of thinking about the original problem as they impose a natural structure on the physical (distinguished) parameter space of the original problem.
But perhaps the most remarkable property that is shared by the versal unfolding construction is that in distinction to the original system which it replaces, versal unfolding \emph{families} are structurally stable, that is contain all stable perturbations of the original equation. In this sense  the versal unfolding system represents in a somewhat exaggerated language, a (or perhaps `the') `theory of everything' of the original equation.

\subsection{Bifurcation diagrams}
A subtler property of the versal unfolding is related to the problem of the \emph{likelihood} of the equations $g(x,\lambda)=0$, that is how to classify the qualitatively different types of these problems (this is also  related to a set of  more general issues usually referred to as the `measure, naturalness problem' in cosmology). One would naively expect an infinite number of them, one for each value of the unfolding vectorial parameter $\mu$  in the versal family $G(x,\lambda,\mu)$  (compare with our discussion of the main problem in Section \ref{central} below).

However, the singularity theory notion of codimension provides an ingenious approach to this problem in that the likelihood of one such equation of a given qualitative type is higher the smaller the codimension of it. The bifurcation diagram - graph of the solution sets of the versal unfolding equation $G(x,\lambda,\mu)=0$ - provides complete qualitative information about the original physical problem in this respect.
This is so because it describes not only the dynamics of the normal form of the gravity problem, $g(x,\lambda)=0$, but also that of all its perturbations through qualitative analysis of the dynamics of the versal unfolding equation.
In a sense, the normal form equation $g(x,\lambda)=0$ is the organizing centre of the dynamics of the original physical problem, since all possible bifurcation diagrams are generated by its versal unfolding.

Intuitively, in  the simplest, one-parameter case we draw the control parameter \(\lambda\) on the horizontal axis and equilibrium values \(x(\lambda)\) on the vertical, tracing the `valley floors' of our landscape: solid lines for stable wells and dashed lines for unstable ridges.  As \(\lambda\) moves, these curves bend, split, or merge; when a branch vanishes or divides at a bifurcation point, one imagines that a marble riding that floor must `jump' to a new stable curve, modeling an abrupt state change.  In a multi-parameter setting \((\lambda_1,\dots,\lambda_k)\), a more elaborate picture of evolution necessarily emerges,  each point in the parameter space carries its own terrain, and the space is tiled into regions (the `strata') where the phase portrait -- and thus the local basin structure -- remains qualitatively the same.  These regions are separated by bifurcation surfaces (codimension 1), curves (codimension 2), or isolated points of higher codimension.  As one slides through a given stratum, the corresponding equilibrium basin deforms smoothly under the knobs' tuning; crossing a boundary forces the marble into a different valley, illustrating how equilibria appear, disappear, or reconnect in higher-dimensional bifurcation diagrams. In effect, the multi-knob parameter space is a patchwork quilt of `constant-portrait' regions.  A multidimensional bifurcation diagram is just the atlas of these patches and their boundaries, showing not only where wells appear or vanish, but how the very rules of the landscape's splitting change as one moves through parameter space.

A bifurcation diagram represents the complete  solution set of the versal unfolding system. It comprises two closely related  parts:
\begin{enumerate}
  \item The parameter space stratified by topological equivalence,  with $\mu_i, i=1,\dots,k$ being the axes. This space is stratified, i.e., partitioned into subregions - maximally connected sets (possibly points) called \emph{strata}, by the bifurcation (boundary) sets of the problem. These sets are smooth submanifolds of $\mathbb{R}^k$.
  \item The qualitatively different phase portraits corresponding to each stratum. The phase portrait at the origin of the bifurcation diagram represents that of the (normal form of the) original problem.
\end{enumerate}
There are consequently various interesting phenomena associated with the bifurcation diagram of any given gravity problem. We note that these phenomena are parameter-dependent and so  disappear as the parameter $\mu\to 0$, that is they are totally absent in any phase analysis of the original problem. For example, one observes the various possible  metamorphoses of any single phase diagram as the parameter point moves around in the stratified space and passes through the various bifurcation sets.

Various examples of gravitational bifurcation diagrams are given in the next Section (cf. \cite{cot23}-\cite{cot25a} for more details).
\subsection{Moduli}
These are further types of parameters other than the distinguished and auxiliary (i.e., unfolding) ones. \emph{Modular coefficients} (or, moduli) appear in the normal form and the versal unfolding equations as integer coefficients in front of some terms. The reason for their appearance is related to the existence of certain perturbations of $g$ in the versal unfolding which are topologically  -  but not $C^\infty$ - equivalent to $g$ itself.
In this sense, one associates a \emph{topological} codimension to the equilibrium of $g$ defined as its codimension minus the number of moduli coefficients.

In singularity and bifurcation theory, the moduli appear sometimes as the intrinsic shape parameters of a landscape that survive all smooth re-coordinations and parameter redefinitions.  In our terrain picture, imagine two adjacent wells whose relative depths or the heights of the ridge between them cannot be equalized by any mere tilting or stretching of the ground; those persistent depth-ratios and ridge-heights are the moduli.  Each choice of moduli fixes a genuinely different `terrain type' that cannot be smoothed into another, so they label inequivalent bifurcation patterns and stratify the full space of possible landscapes. Moduli thus label genuinely different bifurcation diagrams, which cannot  transform into each other by merely re-coordinating variables or reparameterizing our `control knobs'. In other words, two systems with different moduli sit in different strata of the unfolding classification: their diagrams show distinct patterns of branch splitting or merging that no smooth change of variables can reconcile.

The way to calculate the moduli is through algebraic relations that define the notion of genericity for versal unfoldings, these being conditions on the derivatives of $g$ of two kinds: nondegeneracy conditions involving derivatives with respect to the phase variables, and transversality conditions on derivatives with respect to the auxiliary parameters; the former conditions are needed to ensure that the equilibrium of the system is not too degenerate, whereas the latter are required so that the parameters unfold the field in a generic way.

Although as we shall see there are moduli coefficients in the constructions of most of the problems developed in the next Section, we do not use the topological notion of equivalence.

\subsection{Symmetries}
It has probably become clear by now that in dynamical bifurcation theory one is not really interested in seeking an analytical expression for the solution of a given problem. The general attitude is dictated by the motive that since even in those rare cases where something like that becomes feasible, it is not always clear what such a solution represents, and so the interest is now rather shifted to obtaining a complete \emph{topological} information about the global behaviour of the orbits by looking not only at the original problem given but also at nearby ones (in some suitable topology).  In this situation, the presence of \emph{symmetries of the equations} describing the laws \emph{themselves} rather than of their solutions (apart from those already used to reduce the original field  equations to the problem-idealization currently dealt with)  becomes an issue of paramount importance. In particular, one usually allows for  versal unfoldings which respect the symmetries of the original problem.

To define the notion of a symmetric dynamical system, let $\Gamma$ be a compact Lie group and consider its action of the vector space $V$ of the dependent variables $x$ of the problem $\dot{x}=g(x)$. We say that $g$ is $\Gamma$-\emph{equivariant} if for all $\gamma\in\Gamma$, we have $g(\gamma x)=\gamma g(x)$. In  particular, if $x_0$ is an equilibrium of the system, then so is $\gamma x_0$.
In this case, either $\gamma x_0\neq x_0$, and we have found a new equilibrium, or $\gamma x_0=x_0$ and $\gamma$ is a symmetry of the solution $x_0$ (this is useful because we may then enumerate only those equilibria not related by the symmetries of the problem).

There is a $\mathbb{Z}_2$-symmetry in some of the problems employed below, and indeed many other gravitational problems  are equivariant with respect to some symmetry group.
For example, the Wianwright-Hsu equations for homogeneous cosmologies are $D_3$-equivariant where $D_3$ is the dihedral group, and this symmetry has important implications for the early evolution and transient behaviour of Bianchi dynamics (cf. \cite{cot2208}, for a preliminary discussion of this problem).

In general, the  presence of symmetry in a gravity problem may have decisive implications for its  solution. For instance, in the problem of singularities treated in Section \ref{sing}, the symmetry of the equations   has a very important implication for its codimension of that problem: since the linear part of the normal form is the zero matrix, it means that in principle one needs to unfold using four parameters. However, the presence of a  $\mathbb{Z}_2$-symmetry in that problem \emph{reduces} the codimension of two, therefore we only need to unfold the field with only \emph{two} parameters.

The use of equivariant singularity theory \cite{golu3} in gravity problems, as for e.g.,  in the problem of singularities or in the Friedmann-Lema\^itre equations below) becomes  an important factor in the progress that can be made given the very high complexity involved in the original equations.

\subsection{Global bifurcations}
Up till now we have been deploying local bifurcation theory, that is looking for bifurcations in small neighbourhoods of nonhyperbolic equilibria or closed orbits (cf. the bifurcation diagrams). However, bifurcations of a \emph{non-local} character are also possible for gravitational systems and in fact may be a decisive factor.

These are related to changes in the phase portraits in small neighbourhoods of so-called homoclinic and heteroclinic orbits, and global bifurcation theory provides the platform to study such effects in gravity. Homoclinic orbits in phase space connect asymptotically a hyperbolic equilibrium to itself whereas heteroclinic orbits connect in this way two such equilibria, both describe non-local phenomena.

Intuitively, local bifurcations tweak the valley floor near an equilibrium, but \emph{global} bifurcations reshape the large-scale terrain by altering the connections between valleys and ridges.  A homoclinic orbit is like a marble that escapes its home valley by rolling over a saddle ridge, loops around the landscape, and then returns to the very same valley, so that its unstable and stable manifolds intersect.  A heteroclinic orbit instead connects two different valleys: the marble leaves one basin, crosses a ridge, and settles in a neighboring one.  As parameters vary, these global loops or connections can form or break, for instance,  when a ridge lowers just enough to let the marble cross, producing sudden, system-wide changes in dynamics such as the onset of complex or chaotic behavior.

Although such orbits have not been studied in gravity to any systematic degree, they indeed appear abundantly in some of the problems studied so far in gravitational bifurcation theory (for instance as saddle-connections). Also they are expected to appear and  play important roles in bifurcation theory analysis of the partial differential equations that arise when the original problem has an inhomogeneous character (after centre-manifold reduction).

Therefore global bifurcations constitute an inherent feature of gravitational equations and deserve to be further studied. In particular, the typical ways of how and when the breaking of such orbits occurs via homoclinic and heteroclinic bifurcations and new invariant sets are thereby created is currently unknown.

\subsection{Physics of the versal unfoldings}
The versal unfolding and its bifurcation diagrams have non-zero parameters everywhere, except at the origin of the parameter space $\mu=0$ which corresponds  to the original system.  All unfolding parameters are zero in the original system (and its normal form). Therefore the physics associated with versal families is expected to possess characteristics and properties foreign to those  which one is accustomed with when working solely  with the original system. Physically speaking, a versal unfolding then gives us a master family of  `landscapes' (by adding just enough extra parameters to capture every way that flattening can be deformed). Each choice of unfolding parameters picks out one particular landscape, and slicing that landscape along a chosen path in parameter space produces exactly the familiar bifurcation diagrams. In this way, the versal unfolding construction really is the `set of all possible landscapes', and the collection of its bifurcation diagrams tells us, for every way one might vary $\mu$ how the equilibrium sheets appear, merge, or disappear.

Below we assume that we have constructed  a versal family corresponding to some original system (e.g., that found by some metric and matter source in relation to the Einstein equations) by  starting from the latter  and following the theory described in previous subsections.
Then new physics will emerge as a result of the cooperation of the following factors which are only linked to the  versal unfolding.
\begin{enumerate}
\item All versal physical effects  arise due to the unfolding parameters, and are described by  smooth and gradual transitions between phase portraits upon parameter change in the bifurcation diagrams. These effects generally disappear when the unfolding parameters are all set to their zero values.
\item The versal unfolding system is structurally stable.
\item All features of the bifurcation diagrams persist under perturbations.
\item The structural stability of the versal unfolding implies that all solutions are free of spacetime singularities of other divergences of the physical fields.
\item At a bifurcation set defined in parameter space, the system transfigures to a new stage of its evolution described by the phase diagram of the next stratum and becomes qualitatively inequivalent to its previous form.
\item Local bifurcations  lead to an appearance of quasi-global results about the original physical problem.
\end{enumerate}
We shall discuss various aspects of these effects and constructions in the next section that will put known properties of the original gravitation  systems in a new context. We shall see that physical effects associated with the versal unfolding equations rather than the original ones contain much more diverse information than what one is accustomed with in the non-unfolded problem that one has started from. Because of its structural instability, the slightest deformation of the original system (now lying at the origin of the bifurcation diagram) will lead to a series of new and qualitatively inequivalent phase portraits and their continuous changes to other forms upon parameter variation.
As shown in great detail in the bifurcation diagrams of the next Section, that complexity is due to two factors: 1) the parameter space is higher-dimensional (rather than zero-dimensional that is a `point' when the parameters are zero as in the original system), and 2) the centre manifold is also higher-dimensional (than zero).
More importantly, the Jordan form of gravitational problems at the origin is typically  a degenerate matrix (usually with a number of zero eigenvalues).

\section{Gravitational bifurcations}\label{bifningr}
\subsection{The main problem: intuitive discussion}\label{central}
For a gravitating system modelled by a dispersive (i.e., structurally unstable) set of dynamical equations, there are two main, very different qualitatively, cases to consider:
\begin{enumerate}
\item 	Arbitrary small perturbations of the equations give rise to an infinite number of topologically inequivalent types of behaviour,
\item 	Such perturbations give rise to only a finite number of topological types.
\end{enumerate}
This is in sharp contrast to the possible behaviours of structurally stable, hyperbolic systems as discussed in the previous section of this paper. We say that gravitational equations are of infinite codimension in Case 1, and of finite codimension in Case 2. Although one does not rule out the existence of gravity problems belonging to Case 1, in any encountered problem of infinite-codimension one would typically obtain overdetermined problems and following R. Thom \cite{thom}, chap. 3, one expects to be able to show that there exists a finite order such that `almost all' enumerations at some smaller order become determinate at that order. This means that in this case higher-order terms beyond that order may be neglected, and we may treat this case as one of finite codimension\footnote{There is also a more `pragmatic' approach to versal unfoldings, where one truncates the jet at some order $k$ and considers  perturbations with finitely many parameters for which the jet is $k$-determined only with respect to some given property (or properties). The unfolding will then be versal only with respect to those properties, \emph{not} with respect to all. Although this is a very interesting alternative approach to versal unfoldings, it is not the one that is followed here. For more details on this approach to bifurcation theory (and for a number of more advanced/alternative treatments of  topics of the more standard approach followed here), the reader is directed to Ref. \cite{mur}.}.

We can now state the central problem addressed in this paper:

\medskip

\noindent\textbf{(a) Physical formulation:} Determine up to equivalence the number of different topological types of behaviour which arise due to gravitational structural instability.

\medskip

\noindent\textbf{(b) Mathematical formulation:} Determine all possible topologically inequivalent types of forms that are imposed by the defining gravitational equations.

\medskip

The case of finite codimension, say equal to $s\leq 4$, is the main case we shall consider below. In fact, we shall advance a point of view and various results that point to the finite codimension answer, and for various reasons explained below-while we keep an open mind-we also expect that in most cases of interest the codimension for gravitational systems will be small (up to four).

Let us assume for concreteness that we are given a gravitating system governed by the Einstein equations (for example, a Friedmann-Robertson-Walker (FRW) universe with a perfect fluid source-the following argument remains applicable in any modified gravity and effective string theory situation, etc). There are two basic issues on which our approach to gravitational bifurcations focuses: firstly, the question about \emph{finite determinacy}:  to what extent the low-order terms in a Taylor expansion of the gravitational bifurcation problem completely determine its qualitative behaviour regardless of the higher-order terms that may be present? Secondly, the question of \emph{unfolding}: how does a gravitational bifurcation depend on parameters? The natural area to formulate both questions and search for complete answers is provided by the elaborate mathematical framework of bifurcation theory. Armed with the ideas of the previous section, we imagine that when an equilibrium $(x,\mu)=(0,0)$ at the origin of a parametrized family of  systems undergoes a bifurcation at $\mu=0$, then the flow for $\mu$ near zero and $x$ near zero is qualitatively different to that near $x=0$ at $\mu=0$.

We may now describe the main objectives of the present approach in a somewhat more precise language. In the process we shall discover how the main methods of bifurcation theory discussed in Section 2 fit into the general problem of gravitation.

\subsubsection{Objectives}
In any gravitational context for example, cosmology, black holes, modified gravity, or effective string theory, we aim at a complete realization of the program of bifurcation theory for a variety of gravitational problems. This program is one of deciding the following goals:
\begin{enumerate}
  \item Determine the nature and types of bifurcational behaviour and decide on the degree of degeneracy of the problem.
  \item Resolve the problem of higher-order terms and the issue of parameters.
  \item Discover the physical implications of the constructed versal families associated with the problem.
\end{enumerate}
This program will be accomplished below by achieving progress in three directions  that together offer a complete portrait of all possible qualitatively different behaviours for a given gravitational problem (the question of \emph{how} to accomplish these objectives was analysed in Section \ref{method}, while examples of how these objectives have worked out so far will be given below.

\medskip

\noindent\textbf{First objective:} \emph{For typical gravitation problems, demonstrate that they are bifurcation theory problems.}

This objective deals with the first main difficulty to realize the bifurcation theory program, namely, to identify the type of degenerate behaviour present in the given gravity problem and separate it from any remaining hyperbolic one.
We have to demonstrate that the given set of gravity equations indeed constitute a bifurcation problem. In this regard, we need  to solve the following problems:
\begin{itemize}
  \item[O1-1] Show that the gravitational equations given are non-generic and therefore the problem is structurally unstable.
  \item[O1-2] Determine the degree of degeneracy of the given gravity problem (Note: a `hyperbolic’ problem is non-degenerate, i.e., its degree of degeneracy is zero). The lowest nontrivial such degree is one.
  \item[O1-3] Remove any hyperbolic behaviour and reduce the dimensionality of the phase space of the problem.
\end{itemize}
It is accomplished in the following three technical steps:
\begin{itemize}
  \item[1a.] Write the given gravitational equations as a dynamical system in suitable variables.
  \item[1b.] Write the dynamical system from Step-a in a way that its linear part is in Jordan normal form.
  \item[1c.] Find the centre manifold reduction of the equations from step-b and determine the dimensionality of the centre manifold.
\end{itemize}
Once we have completed these steps, we have accomplished the said aims [O1-1, 2, 3].

\medskip
	
\noindent\textbf{Second objective:} \emph{Given the results of Objective-1 for a given gravity problem, provide complete answers to the finite determinacy and unfolding questions.}

This objective deals with the second main difficulty to realize the program of bifurcation theory, namely, with the question of higher-order terms and the problem of parameters. Progress here can only be accomplished by dealing with each of the following issues:
\begin{itemize}
\item[O2-1] 	Simplify the nonlinear part of the problem as much as possible. This step solves the `recognition problem', to determine all possible \emph{normal  forms} of the dynamical equations equivalent to the original ones that have the property of finite determinacy.
\item[O2-2] 	Construct a certain distinguished family of perturbations, the so-called \emph{versal unfolding}, that contains all qualitatively different ones.
Classify all possible behaviour that can occur because of the presence of various kinds of parameters.
\item[O2-3] 	
Construct the \emph{bifurcation diagrams}, that is the solution sets corresponding to the versal unfolding. For each connected region (`stratum') in each bifurcation diagram, describe the phase portrait of the corresponding system. Obtain thus a complete picture of the dynamics of the full set of perturbations as defined by the versal unfolding.
    \end{itemize}
Once this objective is completed, one has a novel system of dynamical equations, the versal unfolding, instead of the original system that we started with before the processes of the first objective took place. We emphasize the dimension of the phase space of the emerging system of dynamical equations in the versal family  is equal to the dimension of the centre manifold  in Step-1c in the first objective. Steps O2-1, O1-2 are developed using methods of singularity theory, while in Step O2-3,  one uses dynamical systems methods for parametrized systems (e.g., fixed `branches' instead of `points').

\medskip

\noindent\textbf{Third objective:}  \emph{Develop the physics of versal unfoldings associated with the original gravity problem.}

While the first two objectives  are almost purely mathematical and lead to the constructions of the topological normal forms and the bifurcation diagrams, in this objective the main question is to discover the physical implications stemming from these constructions as regards the original physical problem.
The importance of this objective naturally leads to new physical effects totally absent in the original system. This clearly follows from the fact that the resulting constructions from the first two objectives are very closely interlocked with, and follow from, the degeneracies of the original system, all of which are by necessity ignored in a hyperbolic treatment.
To motivate this objective, let us start by emphasizing the fact that there are three kinds of parameters in the versal unfolding (and associated bifurcation diagrams):
\begin{enumerate}
	\item The distinguished parameters, which are those that possibly appear in the original system of dynamical equations (for example, the cosmological constant, the fluid parameter, or the mass of a black hole, etc),
\item 	the unfolding (or auxiliary) parameters, which are extra ones that appear in the versal unfolding to monitor the perturbations, and,
\item 	the modal parameters, needed to distinguish between inequivalent bifurcation diagrams.
\end{enumerate}
There are at least three main areas of development of the emerging `physics of versal unfoldings':
\begin{itemize}
\item[O3-1] 	\emph{Physics with unfolding parameters:} Since the unfolding parameters describe all possible perturbations of the original system, the first question is related to the physical importance of each one of them. Having in our disposal all possible phase portraits (Step O2-3), it becomes an interesting problem of immediate importance to discover the principle physical effects associated with them.
\item[O3-2] 	\emph{Physics on the `catastrophe manifold’:} There are certain problems where the states of the system are geometrically described as points in the product of the manifolds of distinguished and unfolding parameters (cf., e.g., the crease flow project below). This leads to new physical interpretations of the bifurcation diagrams and so of the solutions.
\item[O3-3] 	\emph{Physics of the metamorphoses:} What is the gravitational significance of each smooth change, or metamorphosis, of a phase portrait? Each bifurcation diagram has a number of phase portraits, one for each stratum, separated by bifurcation sets that is points, curves, or higher-dimensional surfaces where changes of the phase portraits due to changes in the parameters take place.
\end{itemize}

This completes a somewhat detailed description of the main problem of deployment of the bifurcation nature of gravitational problems.

\subsubsection{Importance}

In general, we may say that a gravitational question can be approached in two general ways, the quantitative and the qualitative. In the former approach, one studies an exact or approximate solution of the gravitational equations at hand, or applies perturbation theory methods, with a purpose to obtaining physical conclusions about the asymptotic behaviour, stability or instability, singularities, or a consideration of the possible behaviours of other important physical quantities that are functions of the given solution and its perturbations. In the latter approach, interest is shifted to qualitative properties of the orbits in a phase portrait representation of the solutions with particular emphasis to the equilibria which play a very important role as the organizing centres of the overall dynamics of the system. The qualitative approach to gravity problems is commonly the hyperbolic one as described above.

However, there are three common properties of gravitational equations describing different physical situations which cannot be adequately taken into account by \emph{any} of the traditional methods used for the study of such equations, and lead to the necessity of a bifurcation theory approach to gravity problems. These properties are:
\begin{enumerate}
\item 	The presence of parameters in the governing equations whose exact values are unknown,
\item 	The degenerate nature of the linear part of the governing equations for certain values of the unknowns and the parameters,
\item 	The qualitative change in the behaviour of solutions of the governing equations through arbitrarily small changes in their right-hand sides.
\end{enumerate}
These properties imply that gravitational equations are of an essentially nonlinear and structurally unstable nature. They also point to further important issues that never arise in other standard approaches to gravity problems: how to best understand the types of bifurcations of phase portraits that may occur upon changes of the parameters. In this case, the \emph{bifurcation sets} of the problem replace various distinctive effects associated with a hyperbolic approach, singularities, divergences, isolated equilibria, etc,  and themselves become the organizing centres of the dynamics.

Hence, a bifurcation theory approach to gravity problems is totally different than other, more traditional, methods of study of gravitational equations and their physical effects The latter include increasing the dimensionality, adding other matter terms, or higher-order terms in the action.

In the present approach, the versal equations do not represent a more complex version of the original equations, but rather a certain \emph{completion} of them as we discuss next.
\subsubsection{Gravitational selection and completion mechanisms}
It follows that a bifurcation theory approach to gravitation  leads to the discovery of novel aspects of gravitation that cannot be found without it, and also provides the means for their further study. It implies possibly observable effects and signatures, definite predictions which would be difficult to capture otherwise but ready to be confirmed or falsified.
This aspect of the present work will also be important in physical cosmology starting from bifurcation results. This is so for the following three main reasons:
\begin{itemize}
\item 	Attention is shifted from the study of the original gravity equations to that of their versal unfoldings.

These equations, one may say, provide a consistent, essentially nonlinear method to derive solutions free of singularities. Every result concerning a non-generic system must be accompanied by the determination of its `codimension’, i.e., the number of parameters to be introduced for which the degeneracy is unremovable, and some indication of the possible bifurcations in the so-constructed family.
\item 	Novel mathematical problems appear by necessity.
Bifurcation theory contains  a purely algebraic approach to gravitational problems in which the resulting equations fully determine the solutions. For example:
\begin{enumerate}
\item 	Determine the bifurcation sets as solutions to the algebraic equations,\\ $f(x,\lambda)=0$.
\item 	Find the normal forms up to equivalence of a given set of gravity equations.
\item 	Find the versal unfoldings and construct their bifurcation diagrams.
\item 	Find the behaviour of nearby solutions to the perturbed equilibria of the versal family upon parameter variation.
\item 	Perturb not only the solution but also the system of dynamical equations itself.
\end{enumerate}	

Armed with the versal unfolding results, the  question now arises as to  how the novel  branches - and not just the solutions sitting at the origin of the parameter space -  are lifted, and by which mechanism  are selected, by the `mother' partial differential equation (for example the Einstein equation, which was `collapsed' to the `original problem' due to a symmetry reduction process as discussed earlier). This also includes the general question of stability (in the PDE sense) and extends the problem to that of determining the  \emph{gravitational selection} mechanism which becomes responsible for eventually choosing which branch to keep and which to leave in the process.

\item Novel physical problems appear which require a deeper study.

The resulting versal equations are consistent in the sense of being structurally stable, finite everywhere and depend on a finite number of parameters. The physical importance of the versal unfoldings is usually to be sought-for outside the original framework and into a new one. The versal solutions have various other novel properties that require a deeper study, for example:
\begin{enumerate}
\item In a cosmological setting they describe neither a `big bang' nor a `steady state' situation but exist `in their present form' only between two bifurcation sets in parameter space.
\item 	What happens to one such state when it reorganizes itself after its last metamorphosis? Will the system ever return to it?
\item 	What is the nature of `evolution in time' vs. `evolution in parameter'? Which is more fundamental?
\end{enumerate}	
\end{itemize}
This is where the importance of a gravitational bifurcation approach lies. It represents the first few steps towards a more comprehensive approach of `gravitational selection and completion' by generic perturbations, and independently, an approach to the problem  posed above. For it provides a reliable path from the large body of known and well-understood results related to a hyperbolic approach, to the description of a consistent (i.e., structurally stable) framework, a new but yet unknown structure sought-for in this approach to general relativity, which contains all `surviving' versal unfoldings of a given original idealization. For example, what are the versal unfoldings and the precise relations between them for the Einstein-Maxwell, Einstein-Dirac, or Wheeler-DeWitt equations?

\subsection{Friedmann universes}
\begin{figure}
     \centering
     \begin{subfigure}[b]{0.4\textwidth}
         \includegraphics[width=\textwidth]{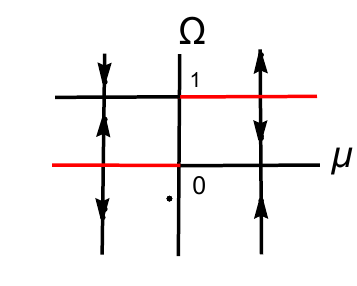}
     \end{subfigure}
     \hfill
     \begin{subfigure}[b]{0.4\textwidth}
         \includegraphics[width=\textwidth]{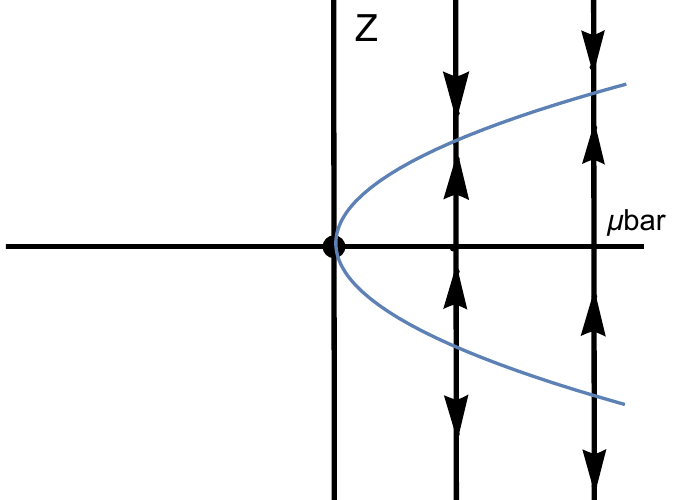}
     \end{subfigure}
                \caption{Comparison of the density parameter evolution on the parametrized centre manifold (i.e., on $H=0$, $\mu$ being a new \emph{dependent} variable) in Friedmann cosmology: standard case (left) vs. bifurcation case (right). On the left we see the phase-parameter diagram of the system (\ref{ooc})-(\ref{m0}), with the $\Omega$ phase spaces being the lines orthogonal to the $\mu$-axis. The  red  half-axes show instability and the horizontal black ones signify stability. On the right we see the cosmological saddle-node bifurcation  diagram of the system (\ref{un3}) for the evolution of the universe. There are two new equilibria -- the stable node and the saddle -- the stable orbit is in the first quadrant. Phase lines are shown for two different $\bar{\mu}$-values.}
        \label{frw}
\end{figure}

In this subsection, we review the main results of the application of bifurcation theory to the Friedmann equations describing simple fluid-filled universes. We follow closely \cite{cot23} and consider the standard Friedmann equations as a 2-dimensional system of evolution equations for the Hubble parameter $H$ and the normalized density $\Omega$, using a dimensionless time variable $\tau$  defined by $dt/d\tau=1/H$,  namely,
\begin{align}
\frac{d\Omega}{d\tau}&=-\mu\Omega+\mu\Omega^2\label{o0}\\
\frac{dH}{d\tau}&=-H-\frac{1}{2}\mu\Omega H.\label{h0}
\end{align}
This system has a number of non-hyperbolic equilibria of which we shall only consider the non-hyperbolic Milne state, namely, the $\Omega$-axis,  EQ-1: $(\Omega,H,\mu)=(\Omega,0,0)$ (for the full analysis of the  dispersive steady states we refer to  \cite{cot23}, Sects. 5-7). It is important to realize that this system is not a true 2-dimensional system near each of the non-hyperbolic equilibria, but the dynamics is consequently restricted to a 1-dimensional invariant manifold along the dispersive dimension. Therefore a centre manifold reduction near EQ-1 gives the following dynamics on the \emph{centre manifold} $W^c_{loc}$ (which in this case turns out to be the $\Omega$-axis)  \cite{cot23},
\begin{align}
\Omega'&=-\mu\Omega+\mu\Omega^2\label{ooc}\\
\mu'&=0.\label{m0}
\end{align}
This analysis leads to Figure \ref{frw} (left)  where  we have sketched the solution set, i.e., the $(\Omega,\mu)$-plane, where $\mu=3\gamma-2$, $\gamma$ being the fluid parameter of a fluid with equation of state $p=(\gamma-1)\rho$.
We then apply the bifurcation sequence steps  of Section \ref{method} and arrive at the versal unfolding, \cite{cot23},
\be\label{un1a}
\frac{d\Omega}{dT}=\Omega(\Omega-1)+\nu,
\ee
where,
\be\label{nu}
\nu=\frac{\sigma}{\mu},
\ee
is the \emph{unfolding-fluid parameter} and we have rescaled the time to $T=\mu\tau$. In terms of the \emph{versally unfolded density} $Z=1/2-\Omega$, that is with $\Omega$ being a solution  of (\ref{un1a}), we arrive at the following form of the versal unfolding (\ref{un1a}),
\be \label{un3}
\frac{dZ}{dT}=\bar{\mu}-Z^2,\quad\bar{\mu}=\frac{1}{4}-\nu.
\ee
This is then  the normal form of a saddle-node bifurcation with parameter $\bar{\mu}$, and bifurcation point at $(Z,\bar{\mu})=(0,0)$, and so includes the effects of  perturbing terms of all other orders. The bifurcation diagram is shown in Figure \ref{frw} (right), where $\bar{\mu}=1/4-\sigma/\mu$, with $\mu\neq 0$, and $\sigma$ is the unfolding parameter which describes physically the closeness of the models to the unperturbed Friedmann universe (this can be made precise).

We note that in this problem both the parameter space and the phase space are one-dimensional, and the bifurcation takes place at the origin of the $(Z,\bar{\mu})$-plane.
This leads to physical effects associated with the appearance of the new branch of fixed points, namely, the parabola shown in the Fig. \ref{frw} (right). We note that these solutions have no counterpart in the standard picture (which only has the $\Omega=0,1$ equilibria), and so several features of the standard cosmological picture are now replaced by novel properties stemming from this Figure.

For example, the equilibrium solutions 0 and 1 of the evolution equation for the density parameter $\Omega$ in standard cosmology are now absent and are replaced by new perturbed equilibria, i.e., pairs of the form $Z_\pm=\pm\sqrt{\bar{\mu}}$. This has various interesting implications.
\subsubsection{Three applications of the versal FRW equations}
First, under parameter variation the novel solutions ((un-)stable with the (minus) plus sign) appear on the parabola for positive values of the parameter (the vertical lines being the $Z$-phase spaces), coalesce at the origin and disappear for negative values of the parameter  (the phase portraits for the saddle-node and not shown presently).
The perturbed equilibria possess  novel properties  (we direct the reader to \cite{cot23} for more details and complete proofs), for instance,  unlike the situation in standard cosmology no singularities exist in this case.

Another novel application of the versally unfolded Friedmann evolution  that is impossible with the original Friedmann dynamics governed by Eq. (\ref{m0}) relates to the question of \emph{synchronization}, cf. \cite{cot23} (for other related references on the general problem of cosmological  sync, we refer to \cite{ba20}, \cite{sync1}, \cite{cot2208}).  This is the general question as to whether one or more causally disconnected Friedmann domains can  sync during their evolution if they were not so synced initially (for a single domain $A$, sync means that if it is initially synced it \emph{remains} so during its evolution). This can be made precise by introducing the \emph{sync function} $\omega(\Omega,\mu)$ which measures differences in the density profiles of Friedmannian domains during evolution, and we say that A domain remains synced during evolution provided $\lim_{\tau\rightarrow\infty}\omega(\Omega|_A,\mu)= 0.$ It is a basic result that although the original Friedmann equations cannot lead to domain synchronization in this dynamical sense, the versal unfolding $Z'=\bar{\mu}-Z^2$ indeed synchronizes the universe asymptotically, either to the future or the past. This in turn leads to a novel solution to the horizon problem, as explained in detail in \cite{cot23}.

Still another application of the versal family (\ref{un3}) is related to the existence of novel accelerating solutions absent in the standard model, cf. \cite{cot25a}. The construction of such solutions is an elaborated process and so we shall presently restrict our discussion to only the first step: What is the physical significance of the unfolding parameter $\sigma$ that appears in the versal unfolding equation (\ref{un1a})and in the $Z_\pm$ equilibria? The latter can be written in terms of the $\Omega$ function, that is,
\be \label{un4}
\Omega_\pm=\frac{1}{2}\left(1\mp\sqrt{1-4\nu}\right),\quad\nu<\frac{1}{4},
\ee
and this implies that all  solutions are stably attracted by $\Omega_+$ in the future, and by $\Omega_-$ in the past, and we find that the $\Omega_+$ equilibrium describes universes that are always open, while the $\Omega_-$ equilibria can be open (resp. closed) according to  $\nu>0$ (resp. $\nu<0$). Writing the versal unfolding equation (\ref{un1a}) in terms of the standard energy density $\rho$, we find that it leads to the form
\be\label{modrho}
\dot{\rho}+3H(\rho+p)=3\sigma H^3,
\ee
which means that a nonzero  $\sigma$ determines an out-of-equilibrium evolution and the non-constancy of entropy, with an entropic force term proportional to $\sigma H^3$. It turns out that this is key to understanding the nature of the versal family (\ref{un3}) and has important implications for the $\Omega_\pm$ steady states.  We note only the following: the deceleration parameter $q=\frac{1}{2}\mu\Omega_+,$ implies that there are accelerating solutions with either increasing or decreasing entropy, namely,  $q<0$ even when $\mu>0$ and the strong energy condition is satisfied and \emph{any} $\sigma>0$, or $<0$. Analogous results hold for the $\Omega_-$ states, cf. \cite{cot25a}. We shall refrain from presenting the complete arguments which require an extensive discussion of various aspects of the bifurcation properties of this model \cite{cot25a}.

\subsection{The Oppenheimer-Snyder black hole}
\begin{figure}
\centering
\includegraphics[width=\textwidth]{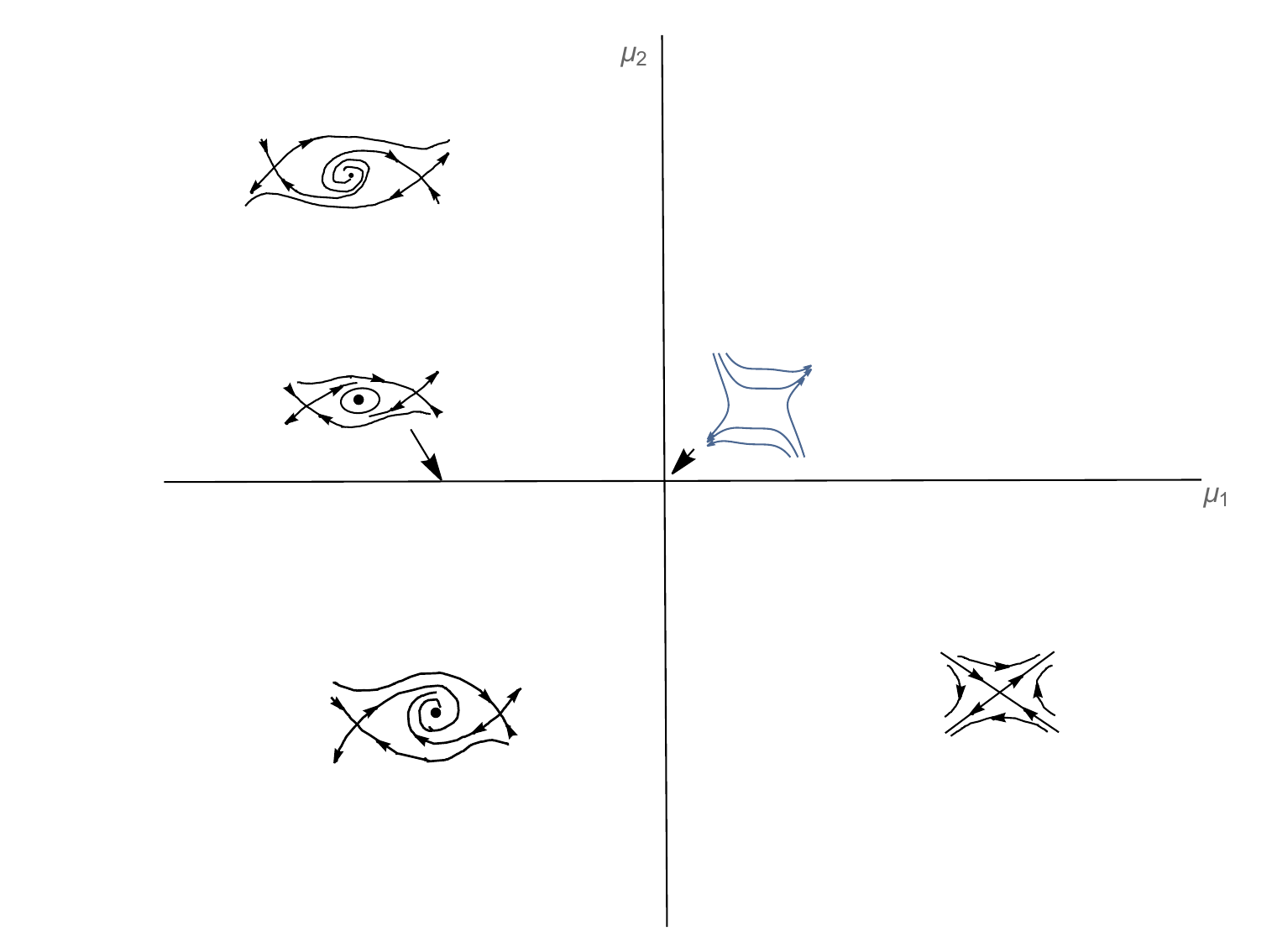}
\caption{The  bifurcation diagram for the Oppenheimer-Snyder-system, positive moduli case, depicting the \emph{centre-manifold(s)-reduced dynamics}. The bifurcations are as follows: there is a pair of pitchfork bifurcations taking the system along the  fragments from 1st to 2nd quadrant, being supercritical  in direction right to left ($\mu_2>0$); and from 4th to 3rd quadrant, being subcritical in direction right to left ($\mu_2<0$). Also there is a  supercritical Hopf  bifurcation dominated by the $y=\dot{x}$ variable taking the system in the $\mu_1<0$ half-space from bottom to top, the bifurcating orbit being stable on the horizontal axis. We note that since the system bifurcates, the number of equilibria changes (as shown in this and the other diagrams), thus leading to new phenomena not having a classical analogue. The escaping orbits  signifying collapsing states although appearing everywhere they are never actually realized, as the system transfigures to the `next' possible state upon parameter variation.  }\label{oppie1}
\end{figure}

\begin{figure}
\centering
\includegraphics[width=\textwidth]{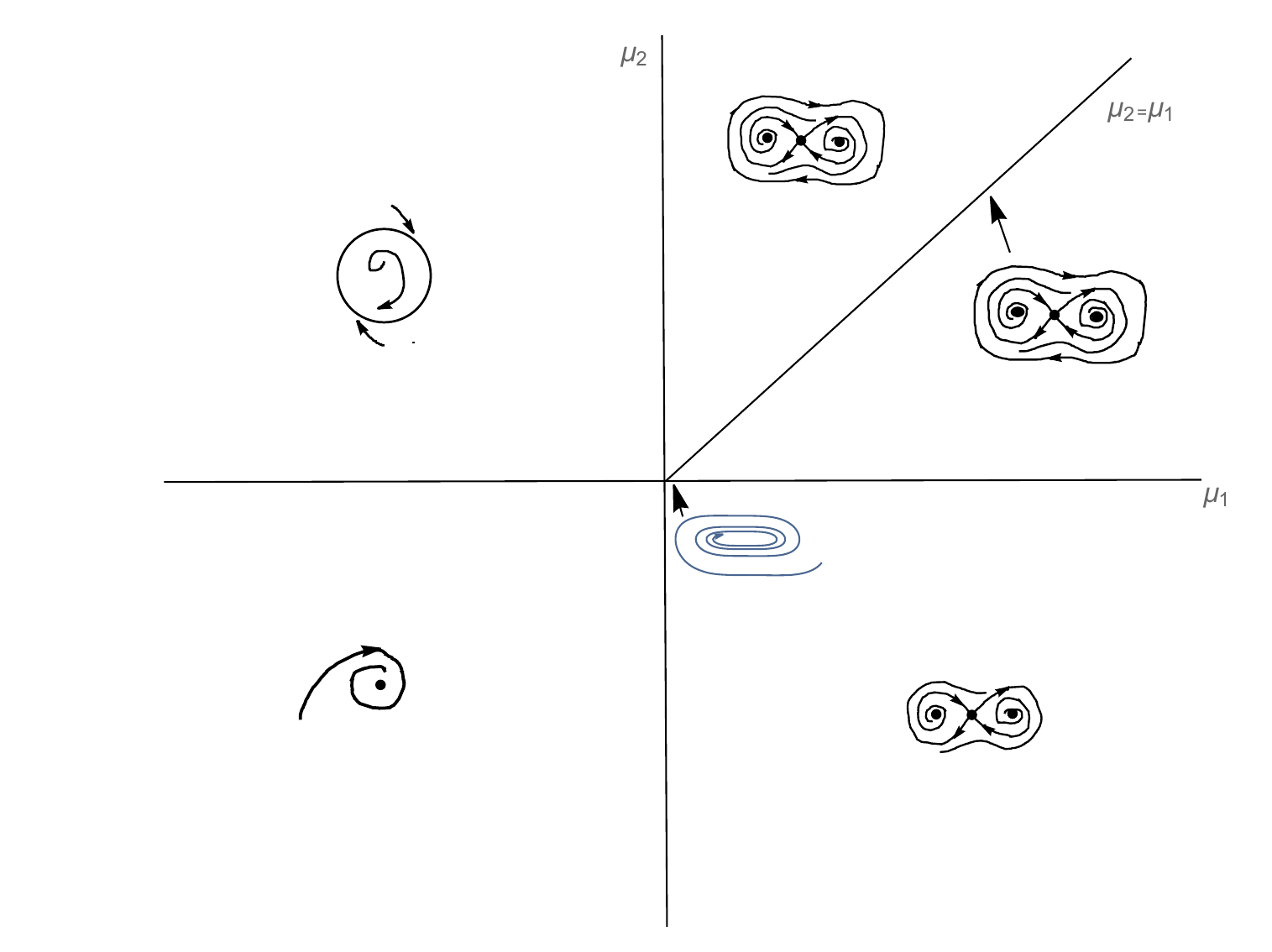}
\caption{The  bifurcation diagram for the Oppenheimer-Snyder-system, negative moduli case, on the centre manifold. This is a somewhat more complicated case that includes also non-local bifurcations described by saddle connections (not shown here). The local bifurcations are described by a) a pair of pitchfork bifurcations from the 1st to the 2nd quadrant, supercritical  in direction right to left ($\mu_2>0$), and from the 4th to the 3rd quadrant, subcritical in direction right to left ($\mu_2<0$); b)  a subcritical Hopf  bifurcation on the $\mu_1=\mu_2$ line, taking the system in the $\mu_1>0$ half-space, the bifurcating orbit being stable on the horizontal axis as before. The solutions are characterized by closed orbits and hence stability, being in a sense closer to describing `defocusing' rather than the focusing states of the positive moduli case in Fig. \ref{oppie1}.  }\label{oppie2}
\end{figure}

In this subsection we closely follow \cite{cot24a}, Sect. 7 and parts of Sect. 8. The standard Oppenheimer-Snyder (`OS') result describing the optical disconnection of the exterior spacetime and the formation of a singularity at the centre of the black hole in a finite time follows from their solution $e^x= (F\tau +G)^{4/3}$, where $x$ stands here for the OS comoving function $\omega$, $F,G$ are arbitrary functions of the radial coordinate $R$, and one uses the Schwarzschild metric in comoving coordinates.
The Einstein equations for this problem are equivalent to the system   \cite{os},
\be\ddot{x}+(3/4)\dot{x}^2=0,\ee for which the phase portrait orbits show the characteristic OS diverging behaviour (cf. \cite{cot24a}, Fig. 12). Further, it is an instructive initial conclusion borne out of an application of the bifurcation/singularity theory steps  described in Section \ref{method}, that \cite{cot24a}:
\begin{itemize}
\item  the OS-system is very degenerate and when written  in normal form the Jordan form is the nilpotent matrix.
\item  the normal form  in terms of the $x,y $ variables (as defined)  includes 3rd -order terms: $\dot{x}=y,\dot{y}=s x^3-x^2y$, with $s=\pm 1$  is a modular coefficient. We shall call this the Bogdanov-Takens normal form of order-three.
\item  all second-order terms in the OS-equation can be eliminated.
\end{itemize}
The versal unfolding for the OS problem thus requires the inclusion of 3rd-order terms for determinacy, and the problem is of codimension two.
We note the \emph{planar} parameter spaces existing here in the OS problem, rather than the parameter lines we met in the Friedmann problem in Fig. \ref{frw}, and the two parameters, namely, deviation from spherical symmetry and rotation, make up the versal unfolding and bifurcation diagrams for the deformations. The latter are shown in the two figures \ref{oppie1}, \ref{oppie2} below corresponding to positive and negative modular coefficients respectively (cf. \cite{cot24a}, Fig. 13, 14).

In these bifurcation diagrams, we imagine a point moving in parameter space on a small circle around the origin and consider the corresponding motion of a phase point moving on some orbit in one of the phase spaces shown. As the parameter point moves around in parameter space, the phase portraits smoothly transfigure to one another, and the phase point drags away from one phase diagram to the next gradually passing through the various \emph{metamorphoses} of the different topological states (i.e., phase portraits in this case) as shown.
Three basic features appearing in these diagrams are:
\begin{enumerate}
\item the initial formation of trapped surfaces corresponding to the phase point being on one of the escaping orbits in the diagram on the right half of Fig. \ref{oppie1}.
\item the formation of stable closed orbits for various values of the parameters in the left part of the  diagram in Fig. \ref{oppie1}, and generally in Fig. \ref{oppie2}.
\item upon parameter variation, the singularities that would have possibly resulted from trapped surface formation and the OS solution are stably bypassed in the present situation. The system instead smoothly transfigures its phase portraits as shown, and \emph{any} phase point is able to follow other orbits thus easily avoiding staying too long on any runaway orbit.
\end{enumerate}
This is a very rich situation with novel features unlike anything in the original OS analysis. We note that there are additional global bifurcations not shown in these figures.

This constitutes a clear prediction of new physics associated with the versal unfolding in this problem. Under parameter variation, a trapped surface may initially form but such a configuration is always unstable and transfigures into something else precisely defined by new states depending on the position of the parameter point in the stratified space (we recall a stratum as a region in parameter space between two consequent bifurcation sets, e.g., between say a bifurcation axis and a bifurcation curve, etc).

While the original Oppenheimer-Snyder study was restricted to a spherically symmetric dust cloud with spatially constant matter density, it has been subsequently generalized to non-constant matter, non-homogeneous dust clouds, scalar fields, etc, with a partial motivation to tackle the cosmic censorship conjecture (see, \cite{ycb09}, Section IV.12, and chap. XIII, for a review of this work with many references to the original sources). The results deployed in this Section may have some relevance to the cosmic censorship problem, and we give here a few brief and very speculative thoughts on this connection.

The use of the word `generic' in this context in the relevant literature in relativity usually refers only to \emph{solutions} (i.e., generic `spacetimes') rather than to the vector field itself (in particular, one considers for instance `generic' initial data sets specified by a 3-metric and extrinsic curvature). Using \emph{vector fields} rather than simply their solutions for this purpose is, however, what is involved in the standard definition of a generic property in dynamical systems.

Therefore the problem becomes one to see what properties can be shared by a `residual subset in the $C^k$-topology' of a set of nearby vector fields, while one usually considers in this context only the `zero-parameter' solutions (that is those sitting at the stratum `origin' of the parameter space in the bifurcation diagram). This is of course directly related to the idea of structural stability as discussed in this work: the notion of `stability of \emph{solutions} under perturbations' is relevant to structural stability and genericity \emph{only} for hyperbolic fixed points and closed orbits (and transversal intersections of stable and unstable manifolds in this case), but not generally for gravitationally unstable or degenerate systems possibly containing non-hyperbolic equilibria etc. Armed with the results described in Figures \ref{oppie1}, \ref{oppie2} one thus expects that in a versal deformation of the problem of cosmic censorship in this context, the possible instabilities and appearance of non-generic naked singularities will probably be replaced by stable configurations without singularities, transfiguring to one another upon parameter variations.
\subsection{Spacetime metamorphoses}\label{sing}
In this problem we discuss evolutionary aspects of the analysis that leads to the singularities in gravitational collapse to a black hole and in a cosmological setting, cf. \cite{cot24a}, Sections 5, 6. In terms of the Komar-Landau-Raychaudhuri and the Newman-Penrose-Sachs equations, this analysis constitutes another manifestation of the nature of implications that are brought by bifurcation theory.

Firstly, it allows for a complete analysis of  nonlinear feedback and coupling effects between the expansion, shear, and vorticity of a geodesic congruence in spacetime. These effects can be  described by taking into full account all degeneracies of this problem. There are two cases,  the convergence-shear and the convergence-vorticity systems (subsystems of the Sachs optical equations), and we shall discuss presently a few of the main ideas and results of the first case (this is in fact the only case considered in the classical references).

\begin{figure}
\centering
\includegraphics[width=0.6\textwidth]{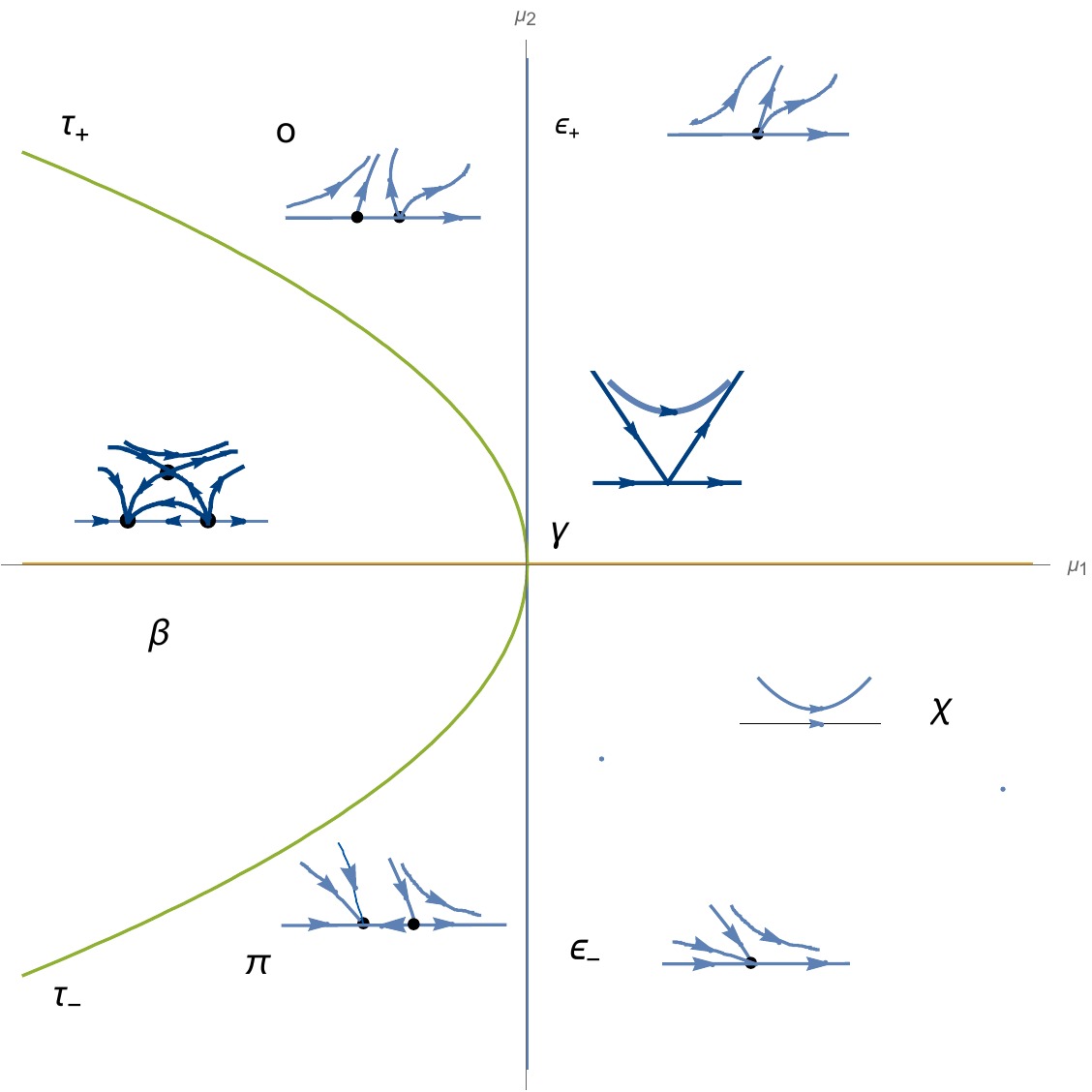}
\caption{The complete bifurcation diagram for the NPR-system (convergence-shear metamorphoses). In this case the centre-manifold-reduced dynamics is governed by the following bifurcations: a) A pair of saddle-node bifurcations dominated by the convergence $\rho$, taking the system along the fragments    $\chi\to\varepsilon_+\to o, \quad \textrm{or, in the opposite direction,}\quad  o\to\varepsilon_+\to\chi,$ and similarly for the negative $\varepsilon_-$; b) A pair of pitchfork bifurcations dominated by the shear $\sigma$, taking the system along the fragments,
$o\to\tau_+\to\beta,\quad \textrm{supercritical, in direction above to below},$ and, $   \pi\to\tau_-\to\beta,\quad \textrm{subcritical, in direction below to above}.$ The $\rho$-dominated, saddle-node bifurcations of the NPR-system take the system from a region of attractive gravity to one where gravity is always repulsive and back, whereas both of  the $\sigma$-dominated, pitchfork bifurcations occur in the repulsive gravity region of the parameter space. According to this picture,  the system in the $\chi$-stratum, or `Hawking-Penrose'-region, of the parameter space will bifurcate through a saddle-node to the strata $o$ or $\pi$ when crossing the positive or negative $\varepsilon$ axis respectively.}\label{bifn}
\end{figure}

\begin{figure}
\centering
\includegraphics[width=0.7\textwidth]{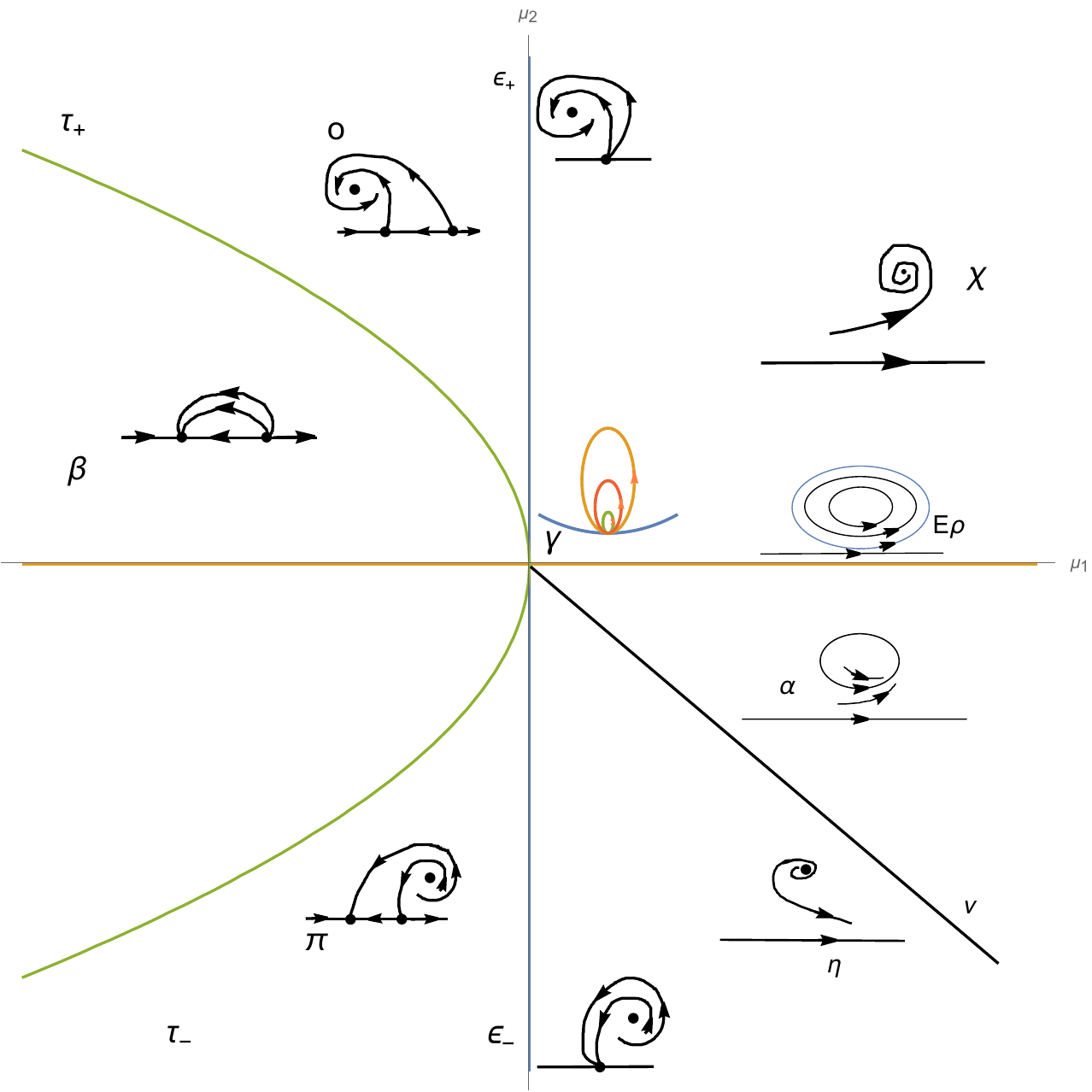}
\caption{The complete bifurcation diagram for the convergence-vorticity system. Here we have again the occurrence of the  saddle-node and the pitchfork bifurcations  similar to those met in Fig. \ref{bifn},  but now with the important difference that a \emph{node} instead of a saddle is involved. Additionally, in this case, mainly due to the combined effect of the convergence and vorticity, there is a third kind of bifurcation, namely, a Hopf  bifurcation dominated by the vorticity $\omega$ taking the system along the  fragments $ \chi\to E\rho\to\alpha\to\nu\to\eta,$ or, in the opposite direction. In particular, the cycle created in the degenerate Hopf bifurcation on the $E\rho$-axis stabilizes on the $\alpha$-stratum and makes the  Hopf bifurcation non-degenerate in the attractive gravity regions  (with the energy condition holding), leading to self-sustained oscillations of the $(\rho,\omega)$ solutions. Upon further parameter variation the cycle disappears to become  an unstable focus in the $\eta$ stratum.  }\label{bifnOmega}
\end{figure}

Secondly, one can now revisit questions concerning  the possible existence of a focussing state. In the classical references this question is of course decided based on the analysis of the Raychaudhuri inequality $\rho'\geq\rho^2$ for the convergence of the spacetime congruence leading to the prediction of spacetime singularities. In the present approach to this problem, we may apply the methodology deployed earlier in Sec. \ref{method} to the convergence-shear and the convergence-vorticity dynamical equations, since these evolutionary models share all those characteristics necessary for a versal analysis. This requires a set of  elaborate techniques which point to the formulation of  codimension-two problems associated with the original degenerate dynamics. The final bifurcation diagrams are shown in Figs. \ref{bifn}, \ref{bifnOmega} respectively. We can only give a few hints here about the global dynamical features revealed by these diagrams. Here  the parameters $\mu_1$ and $\mu_2$ are related to the full projections of the Ricci curvature along a timelike vector and the Weyl curvature along a null tetrad (and rotation parameter  respectively for the vorticity case), and the phase space variables are the convergence and the shear (resp. vorticity) of the geodesic congruence.

As the parameter point moves on a small circle around the origin in the plane, the phase portraits smoothly transfigure to one another, and a phase point rides around from one phase portrait to the next. For instance, the transition from $\chi$  to $\varepsilon_-$ to $\pi$ to $\tau_-$ to $\beta$ strata (or for the opposite direction) in both figures \ref{bifn}, \ref{bifnOmega} is accompanied by metamorphoses of non-equivalent phase diagrams as shown, containing 0 to 1 to 2 to 3 equilibria respectively (or in reverse).

In the emerging physics of the versal unfolding,  standard  conclusions based on the Raychaudhuri inequality  are replaced by several novel physical effects of which we only note here two: the possibility of spacetime metamorphoses, and the `ghost effect' associated with a square root scaling law. Both these effects  point to a forever-existing universe transfiguring from state to state as shown in the bifurcations, and therefore make the existence of an all-encompassing singular state in general relativity a perhaps less plausible conclusion in the sense that it is not a notable feature of the ensuing bifurcations. Had the bifurcation diagram in Fig. \ref{bifn} contained only the $\chi$-stratum (i.e., only the right half-space), there would be one phase portrait and nothing else, and this clearly points to the results associated with the primary case of the use of the Raychaudhuri inequality. In that case, there would be of course no possibility for a metamorphosis, because the focussing state leading to a singularity would be an inevitable feature of the dynamics (corresponding to the diverging orbits in the $\chi$-phase portrait, the only kind of orbit there).

However, here the situation is rather different: the system exhibits two kinds of bifurcations, namely, a pair of saddle-nodes dominated by the convergence $\rho$, and a pair of pitchfork bifurcations dominated by the shear $\sigma$ (recall that while the saddle-node bifurcation is the basic mechanism by which equilibria are created and destroyed, a pitchfork comes in two types, the supercritical by which an unstable equilibrium slowly decays into two new stable ones, and the subcritical where the system destabilizes).
A saddle-node bifurcation takes the system from the region of attractive gravity $\mu_1>0$ where all the energy conditions hold, to the repulsive gravity $\mu_1<0$, and back, while a pitchfork occurs in the repulsive gravity region of the parameter space (supercritical for $o\to\tau_+\to\beta$, subcritical for $\pi\to\tau_-\to\beta$ and back). Hence, the evolution of the system will be mainly characterized by the combined effects of  these bifurcations.
Another notable feature of the bifurcation dynamics, the ghost or `bottleneck' effect, is related to the saddle-node evolution from and to the $\chi$-stratum, the `singularity-forming region'. It implies that the system \emph{delays} its evolution upon entering the $\chi$-stratum with the delay time scaling as $\mu_1^{-1/2}$, a parameter-dependence of the time to the `singularity' as a square-root scaling law.

\subsection{Evolution of event horizons}
The problem of studying crease structures defined as endpoints of at least two horizon generators on the event horizon of a generic black hole as 2-dimensional spacelike submanifolds may be reformulated as an evolution problem of the crease submanifolds governed by a new dynamical system, the crease flow \cite{cot24b}. This replaces the usual Hamiltonian flow  and the crease sets appear as its steady-state (i.e., fixed point) solutions while their bifurcations describe the possible types and topological transformations of the singularities very accurately.

The crease flow introduced in  \cite{cot24b} deals with the dynamics of event horizons at caustic points where the usual Taylor expansions  break down, and consequently it becomes unclear how to make physical sense and describe the evolution that develops such caustics. This approach  has a number of benefits over previous approaches to the problem.
We may write down the defining equations of a wave front for this problem  (obtained from the event horizon by extending the generators beyond their past end points as far as possible) as the union of two null hypersurfaces in spacetime intersecting transversally.
The intersections are given by the equations $u-b_{AB}x^A x^B=0,\,v-c_{AB}x^A x^B=0$, where $(u,v)$ are null coordinates, $A,B=1,2$,  $x,y$, are normal coordinates (the `transverse’ phase space), and we expand in normal coordinates up to second order terms with the $b,c$ being quadratic polynomials in $x,y$.

We can then use suitably defined 2-spheres of radii $\lambda=\sqrt{uv-T^2}$,  $T$ being a time function (synchronous coordinate) in the place of the null coordinates so that to any given point $(u,v)$ there corresponds a value of $\lambda$, and the $\lambda$-lines screen the $(u,v)$-plane. We  then introduce the crease flow for the evolution of the crease sets as that governed by the dynamical system,
\begin{equation}\label{sys2}
\begin{split}
\dot{x}&=\lambda+b_{AB}x^A x^B ,\\
\dot{y}&=\lambda+c_{AB}x^A x^B,
\end{split}
\end{equation}
where the dot means differentiation with respect to the time function, and $x=x^1,y=x^2$ as above. The significance of this system is that in the transverse phase space $(x,y)$, a point $(x_0,y_0)$ is an equilibrium solution of the crease system if and only if it lies on a crease submanifold.

The correct nondegeneracy and transversality conditions on the crease flow imply that the crease evolution is a bifurcation problem of codimension three. This result requires the deployment of several notions of singularity theory into this problem (for instance the nature of the Jacobian of the crease flow at the origin).
Further, one may show that the long bifurcation sequence developed in Section \ref{method} can be applied to the crease flow problem. This leads in particular to three different normal forms and corresponding versal unfoldings (each of which having one distinguished parameter, the $\lambda$, and three unfolding parameters). The results depend on the nature of the zero set and the sign of the discriminant of the Jacobian, making the present problem perhaps the most complicated among all those analysed so far (with the possible exception of the Friedmann-Lema\^itre problem - see below).

We may depict the resulting situation in projections where two of the unfolding parameters are set to zero. In this representation, the problem is described by the bifurcation diagram shown in Fig. \ref{bifnOmega} we met earlier in shifted variables $X,Y$  and parameters $\mu_1,\mu_2$ (one of them is the $\lambda$ and the other one of the unfolding parameters, cf. \cite{cot24b}, Eq. (14)).
In the bifurcation  diagram of Fig. \ref{bifnOmega}, we see the resulting `liquid-like' picture of evolving, in principle observable, bifurcating caustics on the event horizon corresponding to intersections of null hypersurfaces in spacetime. This is based on the joined effects of the three possible bifurcations in this problem, namely, saddle-node when crossing any of the two axes, pitchfork when crossing the parabola, and Hopf when crossing the positive $\mu_1$-axis, acting on the following caustics: swallowtail, folded Whitney, and a pair of `embedded in each other' (i.e., combined) Whitney singularities, all three being steady states of the crease flow. Using  $z$ as a third coordinate, the latter are given by the forms ~$x^2  = zy^2$ (standard Whitney), and $x^2  = z^3  y^2$ (folded).

As the parameter changes differently at different spacetime points, these effects occur simultaneously, thus offering a very rich picture for the evolution of the crease sets. We also note the formation of the stable isolated limit cycle in the $\alpha$-stratum of Fig. \ref{bifnOmega} attracting nearby orbits and disappearing upon crossing the $\nu$-line into the $\eta$-stratum.

\subsection{The Friedmann-Lema\^itre equations}
We shall write the standard Friedmann-Lema\^itre equations in the following equivalent form as a dynamical system for the Hubble parameter $H$ and the fluid density $\rho$,
\beq
\dot{H}&=&-\frac{3\gamma-2}{6}\rho+\frac{\Lambda}{3}-H^2 \label{ds1}\\
\dot{\rho}&=&-3\gamma H\rho, \label{ds2}
\eeq
with the algebraic constraint,
\be
a^2=\frac{3k}{\rho+\Lambda-3H^2}, \label{ds3}
\ee
where the scale factor $a(t)$ is a function of the proper time $t$,  we have a perfect fluid source with density $\rho$ and pressure $p$ with fluid parameter defined by the equation $p=(\gamma-1)\rho$, a cosmological constant $\Lambda$, and we have set $8\pi G=c=1$, while $k=0,\pm 1$ denotes the normalized constant curvature of the spatial 3-slices.

One may show that the roles of the two parameters of the problem $\gamma, \Lambda$ is different: although $\gamma=0$ is a singularity, $\Lambda$ is a true bifurcation parameter with bifurcation point at $\Lambda=0$. In particular, after moving the de Sitter space and Einstein static universe to the origin, split off the linear parts, and set $\Lambda=0$ the Friedmann-Lema\^itre equations take the form,
\be \label{deg-sys}
\left(
  \begin{array}{c}
    \dot{H} \\
    \dot{\rho} \\
  \end{array}
\right)
=-\frac{(3\gamma-2)}{6}
\left(
  \begin{array}{cc}
    0 & 1\\
    0 & 0 \\
  \end{array}
\right)
\left(
  \begin{array}{c}
    H \\
    \rho \\
  \end{array}
\right)
+
\left(
  \begin{array}{c}
    -H^2\\
-3\gamma H\rho \\
  \end{array}
\right).
\ee
This is the basic `degenerate core' of the Friedmann-Lema\^itre equations. It is $(\mathbb{Z}_2 +t)$-\emph{equivariant}, meaning that the system  is invariant under the symmetry,
\be\label{t-symm}
H\to -H,\quad \rho\to \rho,\quad t\to -t.
\ee
We shall call the appearance of  this symmetry  in the final topological normal form a \emph{`time-symmetric case'}, whereas  the case without the $t\to-t$ part a \emph{`time-asymmetric case'}.

\begin{figure}
\centering
\includegraphics[width=\textwidth]{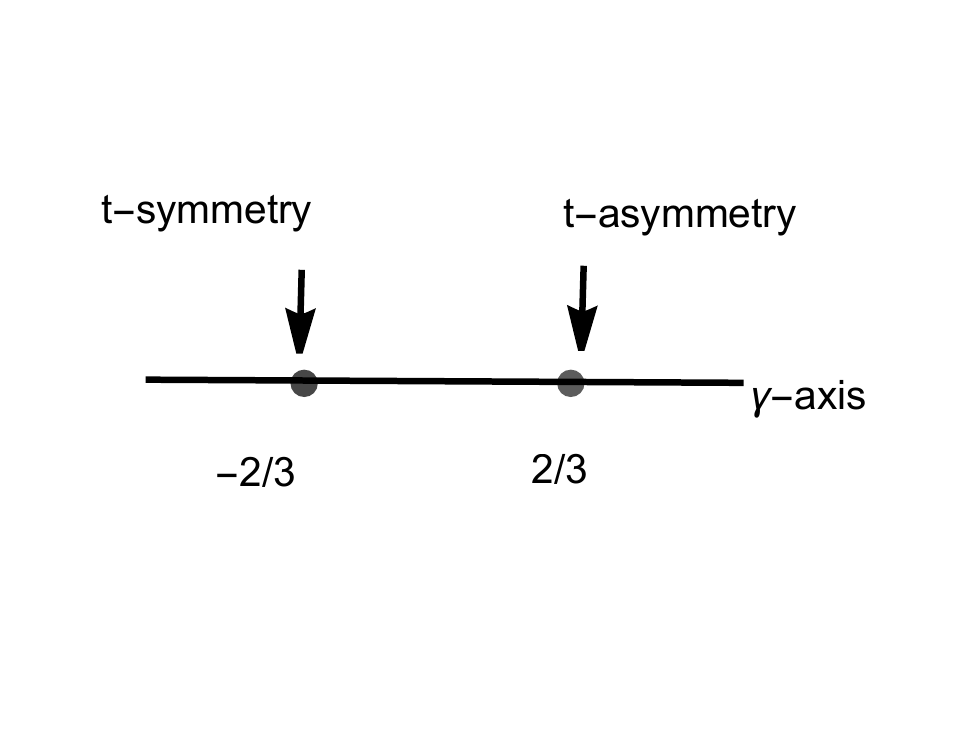}
\caption{The axis of the  parameter $\gamma$ which is responsible for the amazing variety of behaviours stemming from the FL equations as $\gamma$ moves along this axis and jumps from one bifurcation diagram to the next passing gradually from all nine possible such diagrams determined by the $\Lambda$ and the various unfolding parameters necessary in each  case. At the two $\gamma$ values $\pm2/3$, the bifurcation diagrams are qualitatively equivalent to  those shown previously in Figs. 2-5. Away from the two shown values of $\gamma$, the system can be either time-symmetric (i.e., the `FL cubic') or time-asymmetric (i.e., the `FL cusp'). In either case, a passage through the time-(a)symmetric value $-2/3$ (resp. $2/3$) dualizes (i.e., flips)  the behaviour of a time-(a)symmetric that existed for the system when $\gamma\neq\pm 2/3$ to the opposite one.}\label{t-gamma}
\end{figure}

The presence or absence of the time symmetry together with the set of $\gamma$-values shown in Fig. \ref{t-gamma} fully determine the versal dynamics of the Eq. (\ref{deg-sys}) and thus of the original Friedmann-Lema\^itre equations. Following the bifurcation sequence of steps explained in Section \ref{method}, we arrive and the normal forms, versal unfoldings and bifurcation diagrams corresponding to the Friedmann-Lema\^itre dynamical equations.

There are four inequivalent cases for the versally unfolded Friedmann-Lema\^itre system, in total nine bifurcation diagrams. We briefly review this problem in the next four subsections below (for more details the reader is directly to \cite{cot24c}).

\subsubsection{The $\gamma=-2/3$ case}\label{type-3}
The bifurcation diagrams correspond to the so-called cubic Bogdanov-Takens bifurcation, namely,
\begin{equation}\label{va-3rd-A}
\begin{split}
\dot{X}&=Y,\\
\dot{Y}&=\mu_1 X+\mu_2 Y\pm X^3-X^2Y,
\end{split}
\end{equation}
and can be found  in  figs. \ref{oppie1}, \ref{oppie2} for the $(+)$ and the  $(-)$ case respectively. The most characteristic metamorphoses here are related to pitchfork and Hopf bifurcations.

\subsubsection{The $\gamma=2/3$ case}\label{type-4}
The bifurcation diagrams correspond to those of the $\mathbb{Z}_2$-symmetric versal family \cite{zol84}, namely,
\begin{equation}\label{va-2/3-A}
\begin{split}
\dot{X}&=\mu_1+X^2+sY^2 ,\\
\dot{Y}&=\mu_2 Y+ 2XY +X^2Y,
\end{split}
\end{equation}
and can be found  in Figs. \ref{bifn}, \ref{bifnOmega} for the $(+)$- and $(-)$-case respectively ($s=\pm 1$).

\subsubsection{The FL cusp}\label{type-1}
\begin{figure}
\centering
\includegraphics[width=0.7\textwidth]{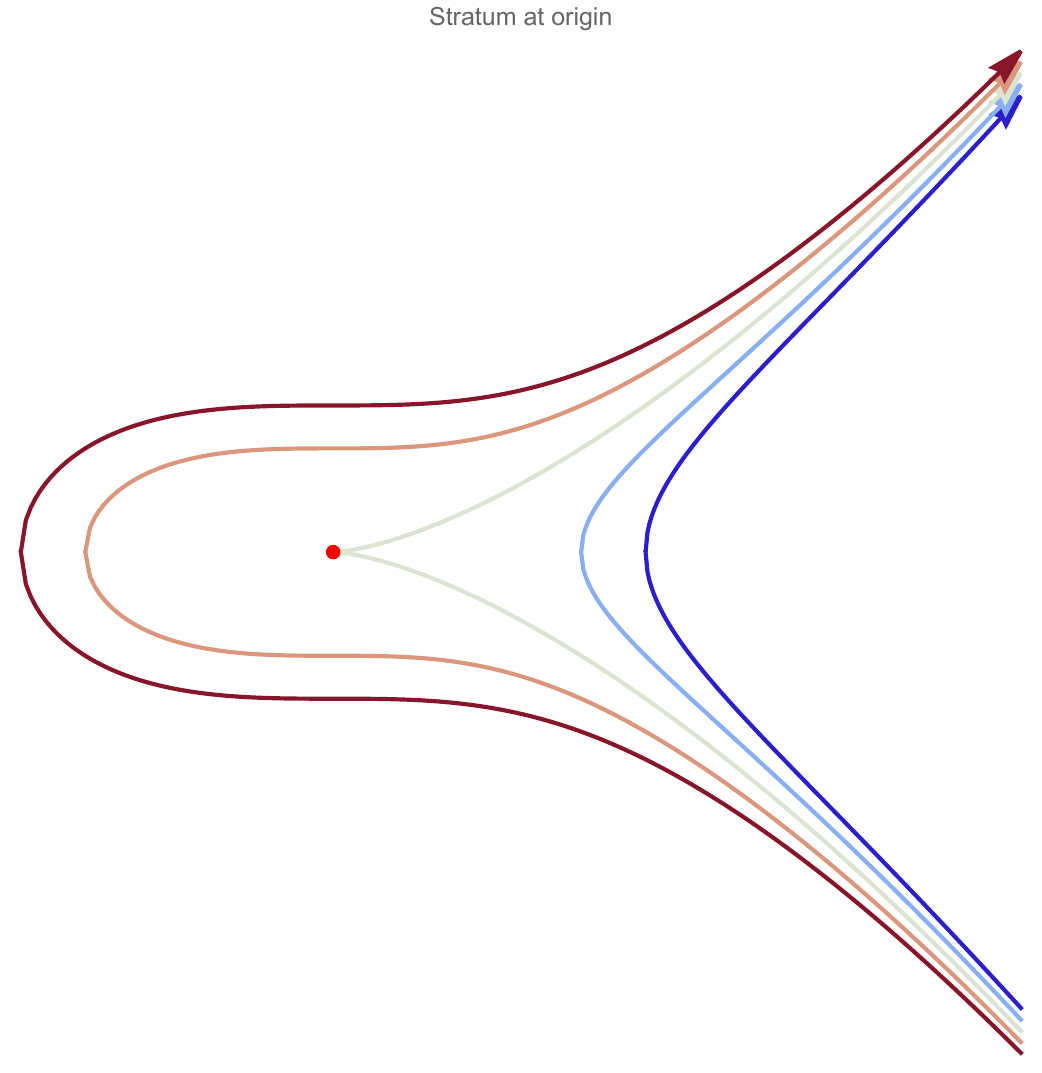}
\caption{The cusp-like phase portrait for the zero parameter limit of the versal family (\ref{va-tot-A}). While a general feature of the quadratic Bogdanov-Takens bifurcation involved in the present FL cusp case is the sequence: saddle-node to  Hopf (cycle creation) to  homoclinic (cycle destruction), the shown cusp-like phase portrait corresponds to the simplest stratum at the origin: Two special trajectories (the stable and unstable separatrices) form the pointed `V'-shape of the semi-cubical parabola; inside that cusp region trajectories slow down as they approach the origin along one branch, linger near the singularity, then are carried away along the other branch. Outside the cusp, orbits simply sweep past, curving around the outside of the cusp separatrices and escaping to or arriving from infinity. The shown cusp-like features are new and correspond to the normal form of the problem at the origin (qualitatively different from the original FL equation).}\label{cusp-case}
\end{figure}

Using similar methods as before, one is led to the construction of the bifurcation diagrams in this case too and the resulting versal dynamics. The bifurcation diagrams  describe \emph{the FL cusp} and  correspond to the \emph{quadratic} Bogdanov-Takens versal family \cite{cot24c},
\begin{equation}\label{va-tot-A}
    \begin{split}
    \dot{X}&=Y,\\
    \dot{Y}&=\mu_1+\mu_2Y+X^2\pm XY,
    \end{split}
\end{equation}
cf. Figs. 6, 7 of \cite{cot24c}. (They may also  be found in, for instance, \cite{wig}, pp. 444-5, for the $(+)$-case, and \cite{kuz}, p. 358 or \cite{du1}, p. 1038, Fig. 15 for $(-)$-case.) For example, the phase portrait corresponding to the origin of the bifurcation diagram is shown in Fig. \ref{cusp-case} (this is the zero-parameter case, cf. Ref. \cite{cot24c}).
\subsubsection{Codimension-3}\label{type-2}
\begin{figure}
\centering
\includegraphics[width=0.6\textwidth]{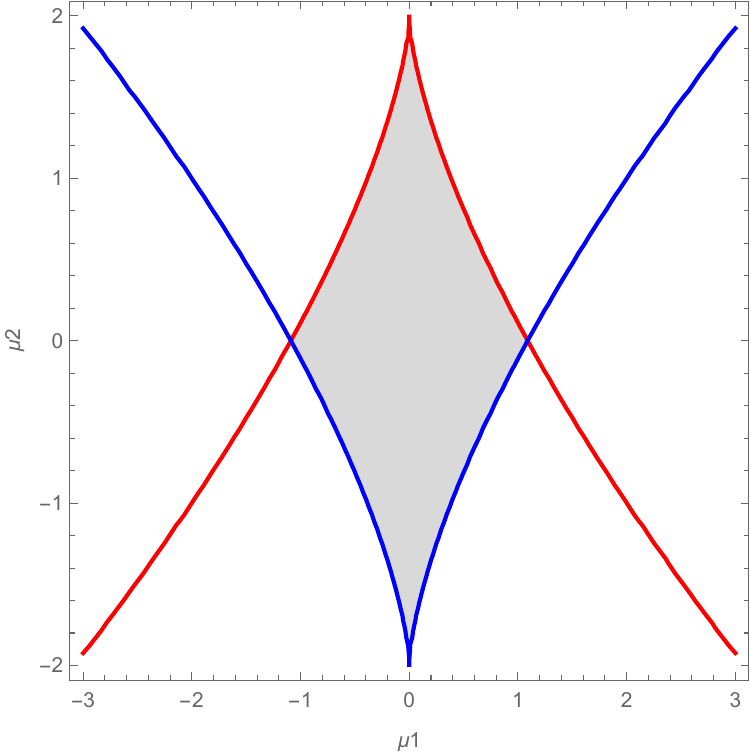}
\caption{The grey area depicts the rhombic structure of the parameter space partitioning it into an internal and an external region for the codimension-3 problem (\ref{va-tot4-A}). The bifurcation diagrams in all cases  are very complex and dependent on specific features that appear in each of the  three cases of the problem. But there is a common theme in all three cases, namely, the existence of the cuspidal `saddle-node' curves shown present in the rhombic region: the condition  for the equilibria of the system (\ref{va-tot4-A}) is the cubic equation,  $\mu_1+\mu_2 x \pm x^3=0$. For the $(+)$-case, the discriminant is  $\Delta_+=-4\mu_2^3 -27\mu_1^2$, while $\Delta_-=4\mu_2^3 -27\mu_1^2$ for the $(-)$-case in the cubic equation. The two bifurcation sets, $\Delta_\pm=0$, for the two pairs of cusps, stratify the parameter space into an \emph{internal region} where there are three real roots to \emph{both} cubic  equations (and so in all three cases: saddle, focus, and elliptic),  and an \emph{external region} otherwise. The gray `vertical-lips' region in between is exactly the set of $(\mu_1,\mu_2)$ where both cubics have three real roots (i.e. $\Delta_+\leq 0$ and $\Delta_-\geq 0$), and form the  cusp-shaped, `rhombic',  region in parameter space shown in this figure.  }\label{rhomb}
\end{figure}

The bifurcation diagrams for the codimension-3 FL cubic (cf. \cite{cot24c}), namely,
\begin{equation}\label{va-tot4-A}
    \begin{split}
    \dot{X}&=Y,\\
    \dot{Y}&=\mu_1+\mu_2 X+\mu_3Y \pm X^3 +B XY,
    \end{split}
\end{equation}
can be found  in \cite{du1}, figs. 3, 4 for the saddle case, figs. 7, 8 for the focus case, and figs. 13, 14 for the elliptic case (see also \cite{du2}). The bifurcation diagram has a rhombic structure with `vertical lips' forming two intersecting cusps as in Fig.  \ref{rhomb}.
This is the dynamically richest case among all versally unfolded FL cosmologies (cf. \cite{cot24c}, and the references given earlier for more details).

\subsubsection{Some remarks}
The relations of the $X,Y$ variables to the original ones for each of the four main case discussed in the previous four subsections is given in \cite{cot24c}. The most important conclusion is related to the global behaviour of the system as $\gamma$ varies, the time-symmetry appears and disappears  as $\gamma$ passes through the two values $\mp 2/3$, and the unfolding parameters also change.

Suppose that  the system is found itself in one of the five possible `ranges' of $\gamma$-values as in Fig. \ref{t-gamma} (the two point values are also regarded as two of these ranges). Then the consequent behaviour of the system is completely determined by the unfolding parameters $\mu_i,i=1,2$ (and also $\mu_3$ for the codimension-3 unfolding (\ref{va-tot4-A})), in the sense that it moves from one bifurcation diagram to the next as $\gamma$ varies, and from phase portrait to the next as the unfolding parameters $\mu$ vary, all these transfigurations being smooth.

We note however that the relations of the versal variables $X,Y$ to the parameters $\gamma, \mu$ are not generally smooth and contain a degree of roughness. This is due to the fact that such inversions are not smooth (or even continuous) because of the violation of the implicit function theorem. Thus although the global metamorphoses of the universes versally are all smooth, this is not true for the relations of the solutions to the parameters of the problem.

\subsection{Some open problems}\label{open}
The following represents a slice of some open research directions and projects to be followed once the reader feels at home with the material discussed in previous sections of this paper. These projects will deploy and/or use the results of the application of the bifurcation sequence of Section \ref{method}. This is not meant to  be an end in itself but rather a means to solve physical problems and approach conjectures that depend on (or arise from) an initial treatment of some set of governing gravitational equations. Gravitational bifurcation problems are also beneficial for the \emph{bifurcation} methods themselves because of the essentially nonlinear and fundamentally different nature of the gravitational equations as compared to other areas of theoretical physics.

\begin{enumerate}	
\item Physical effects for Friedmann-Lema\^{i}tre cosmologies and their metamorphoses.

The extension of Friedmann bifurcations to models with a cosmological constant as discussed earlier presents certain subtle issues which have no counterpart in the simplest models. The main change from a bifurcation theory point of view is that the problem becomes of codimension three (rather than one) and that the cosmological constant is the true bifurcation parameter (rather than the fluid parameter). There are various physical consequences of this result, especially those related to the FL cubic of codimension three and the purpose of this project is to substantiate and further elaborate on certain physical implications that have no counterpart in the original approach to the FL equations and their solutions.

\item 	Cosmic acceleration and Friedmann cosmology.

The new Friedmann solutions shown in Fig. 1b  lead to accelerating solutions that satisfy all energy conditions. This result may be applied to various problems that are usually attacked through dark energy in an effort to see if they could represent a reliable alternative.

\item	Metamorphoses and singularities.

The physics of the versal unfoldings of singularities and black holes as in Fig. 4 may be developed to explore models which rely on evolution through metamorphoses rather than  on the barrier of spacetime singularities. A qualitative comparison of these two aspects of evolution of gravitating systems may be explored in a variety of contexts.

\item 	Relativistic causality and versality.

Causal structure techniques may be deployed to describe generic aspects of spacetime bifurcations. These include the development of `parametric causal structure theory', for instance, the parameter dependence of the boundary of the future of a trapped surface during the crossing of a bifurcation set, as in the bifurcation evolution $\chi\to\epsilon_+\to o$ in Fig. 4.

\item 	Acausal versal null hypersurfaces.

The crease flow evolution may be calculated for a variety of event horizons and its acausal character may be checked in all cases. A very detailed analysis of the physics of all versal unfoldings could then  be carried out.

\item 	Massive scalar fields, the measure problem, and the versal extension.

The massive scalar field model used in the construction of a measure in the set of all universes may now be revisited. This represents mathematically an exemplary problem in the so-called \emph{Hopf-steady state} bifurcation. The possibility of a finite, parameter-dependent, measure for the Gibbons-Hawking-Stewart construction and related works may then  be fully explored.

\item 	Transfigurations in scalar field cosmologies with exponential potentials.

The highly nontrivial mathematical problem of a homogeneous and isotropic scalar field with an exponential potential (which has been studied in many papers) may be explored alternatively by applying the bifurcation sequence, in an effort to discuss the effects of degeneracies on a number of related issues such as the inflationary attractor, the existence of singularities, the nature of initial states, the existence and importance of scaling solutions, as well as various extensions to multiple scalar fields.

\item 	Bifurcations of Bianchi cosmologies of class A.

The Wainwright-Hsu equations, which form the basis of many studies in Bianchi models, possess a kind of dihedral symmetry which makes the deployment of bifurcation methods very interesting (cf. \cite{cot2208}). The problem of the description of Bianchi models through their metamorphoses has not been addressed to any systematic degree so far. We expect to explore a number of fundamental attributes of anisotropic cosmology from this or similar viewpoints which will hopefully shed new light to a number of basic questions.

\item 	Hamiltonian cosmological bifurcations.

There are many problems in cosmology which admit a Hamiltonian formulation, especially anisotropic models. The theory of Hamiltonian normal forms is of course a time-honoured subject, but to the best of our knowledge has not been applied in this framework before, and the same is true for Hamiltonian bifurcations. There are a number of interesting problems in the framework which may be studied in an effort to address some of the many conjectures in this field.

\item 	Cosmological perturbations and bifurcation theory.

Using covariant methods in conjunction with, for instance, the Wainwright-Hsu equations, one is able to write down autonomous systems for many cosmological density perturbations problems. Although the stability of some cases has been studied, the methods are solely based on hyperbolic analysis. A program to revisit this whole area and apply the bifurcation sequence in an effort to decide which of the conclusions of this field remain true in this more general case may therefore be worthwhile. This will lead to a new approach to some of the problems of cosmological structure formation.

\item 	Bifurcations of Brans-Dicke cosmologies.

The purpose here is to find the complete spectrum of bifurcations of Brans-Dicke cosmology for FRW models with a $\gamma$-law perfect fluid with or without a cosmological constant. The hyperbolic analysis has been done in various papers, but this project may also serve as an introduction to more complicated problems of a similar nature that arise in the context of effective string cosmology.

\item 	Bifurcation analysis of string cosmologies.

One may think to  revisit all five effective string theories and present qualitative analyses of flat FRW and Bianchi I cosmology in an effort to determine the nature of their different unfoldings. This analysis may delineate from another point of view the range dynamical behaviour of all five theories in these contexts and help to compare their predictions in concrete situations. Once one has this result, a  study of the effects of other matter fields and spacetime geometries to better understand the nature of such predictions may thus be feasible.

\item 	Bifurcation theory of M-theoretic cosmology.

This project studies 11-dimensional supergravity cosmologies compactified to 4-dimensional universes of various forms. For flat or open FRW universes a hyperbolic analysis has been done. A plan to apply in full the bifurcation sequence to a number of different 4-dimenisonal cosmologies in an M-theoretic context and compare the results to those of the string theory project may then be an interesting one.

\item 	Branes and bifurcations.

We may reconsider the bulk-brane problem for 5-dimensional bulks with embedded de Sitter, anti-de Sitter or flat branes, for polytropic bulk fluids with or without a cosmological constant, taking into account the degenerate nature of the governing equations. Once one has constructed the possible bifurcation diagrams for this problem,   properties of gravity localizing, bifurcating solutions on the brane which satisfy all energy conditions may be another project along these lines.
\end{enumerate}

The last three projects are of a somewhat different nature than the previous ones. All projects discussed above may be considered to be in a sense a necessary step for this purpose.
\begin{enumerate}
\item 	The bifurcation sequence for the Schr\"{o}dinger, Maxwell, and Dirac equations.

Starting with calculations of the zero manifolds of the principal symbol in various cases like in vacuum and certain homogeneous media, one is  hopefully led to a bifurcation treatment of the Schr\"{o}dinger equation, the (linear) Maxwell or Dirac equations, and to further studies of topological invariants (like the Maslov index) that emerge in wave propagation in such contexts, and effects coming from the passage of rays through caustics.

\item 	Bifurcation theory with the full Einstein equations.

It is a standing project to treat the full Einstein equations as a bifurcation problem and determine their (finite) codimension. The first step is to study the symbol of the Einstein equations and determine the zero manifold for a number of cases (vacuum and with matter). This is basically a problem of an algebraic-geometric nature.

\item 	Towards a (new?) fundamental theory.

The versal unfolding equations modify the original set of gravity equations by adding new parameters and extra terms. Why is this modification physical rather than a contrived one? There is a `paradox' involved here because the nonhyperbolic equilibria necessarily present in the original theory (say, general relativity), cannot really be treated properly unless one goes outside that theory and into one that is apparently qualitatively different than others and described by the versal unfolding construction.

In other words, if one insists on studying any original theory per se, one must restrict their studies to only the `hyperbolic' features of it. On the other hand, the versal unfolding of any equation (say the Friedmann equation, or that of the OS problem) exhausts in a sense everything that is to know about it, it represents a kind of a `theory of everything' of that equation. Since all of the original equations belong to one and the same theory as special cases/problems of its full equations, the same must be true for their versal unfoldings. As discussed above, this seems to be in good correspondence with the principles set out in Ref. \cite{ein}, p. 166. What is the nature of the structure that describes all these versal unfoldings\footnote{We note that this interpretation of gravitational versal unfoldings is inequivalent to the one given here, say, for the versal unfolding of the simplest Friedmann equations, as an entropic effect and far-from equilibrium evolution. This is basically an interpretation \emph{within} the framework of general relativity. On the contrary, interpreting the set of all versal unfoldings as comprising a (possibly) consistent theory  is something completely different.}?
\end{enumerate}

\section{Discussion}
This work describes  the path from the large body of known and well-understood results related to  hyperbolic-like approaches to symmetric reductions in the form of differential dynamical systems  of the Einstein equations in general relativity, to  a consistent  structurally stable, `versal'  framework, a new but yet largely physically unchartered structure that emerges when unfolding general relativistic reduced systems such as the five problems described in this paper. 
Our approach to this problem is based on  fundamental mathematical works in the subject of bifurcation theory, starting with H. Poincar\'e's Thesis in the year 1879,  developed further  by A. A. Andronov, R. Thom, V. I. Arnold,  later by M. Golubitski, I. Stewart, F. Takens, R. I. Bogdanov, and more recently by F. Dumortier, H. Zoladek, and many other  mathematicians. In this connection, we  make the further  following remarks.

Consider the Einstein equations with a cosmological constant $\Lambda$, thought of  as a \emph{parameter} in the sense advanced in this paper. This is clearly an unfolding of the vacuum Einstein equations, but is it a \emph{versal} unfolding of them\footnote{An answer to this question could perhaps replace the question mark at the end of the vacuum equations in a famous photo!}?  We have shown that the Einstein equation with $\Lambda$ contains symmetric subsystems, e.g., the FL equations (with $\gamma=0$ in this case), for which the versal unfoldings have codimension greater than one, and  three at most. The same is true for the evolution problem of event horizons, the crease flow also comes with codimension three when versally unfolded. There are further open questions related to these results, both mathematical as well as physical ones lying, however, outside the scope of the present paper and  are presently at the level of pure speculation. A major problem in this direction is that we currently do not have any results on how the unfolded branches can be lifted by the full partial differential equation (we only know this in the zero-parameter case, and there only for the PDE-stability part of this question, not for the conjectured `centre manifold concentration', which in fact may be false).
For the physical direction, one may envisage the set of all versal unfoldings having a particular property, for example  corresponding to homogeneous cosmologies. What is the nature of this set, and what is its physical significance? This set cannot correspond to a kind of modified gravity theory cosmology with additional action terms, because of the following observation.
Consider the FRW problem discussed earlier in Section \ref{bifningr}. In that problem, we start from the well-known results about the behaviour of the simplest FRW models and end up with their versal unfolding, namely, the equation $\Omega'=\Omega(\Omega-1)+\nu$, where $\nu$ is the unfolding parameter. This is a perturbation of the Friedmann equations via adding \emph{lower-order} terms (lower with respect to the first nonzero term in the Taylor expansion), and it is versal i.e., all terms of $O(3)$ or higher have no effect on the nature of the bifurcations. As we discussed, the solutions coming from the versal unfolding equation have entropy-related properties not met in the standard theory and emerge only as bifurcations. In this sense they represent not modifications of the standard theory but a certain completion of it at a deeper level (as we already discussed earlier), one that is very difficult to capture without bifurcation theory. This way gravitational bifurcation theory may considerably expand the present horizon of gravity research.

Obviously one may consider unfolding a theory that contains higher-order curvature terms, or one that is of higher-dimensionality than four, for example in the framework of some modified gravity theory of effective string theory. Although this would lead to  interesting projects in gravitational bifurcation theory, we note that there are already some answers to such a problem stemming from the results discussed in this paper. For example, suppose that one considers a Friedmann universe with a perfect fluid with or without a $\Lambda$ in a modified gravity or string theoretic context. If the linear part of the resulting  equations turns out to be the same as the corresponding ones discussed here, then the answer will be identical to the ones we found here. This will be so due to the power of the versal unfolding construction. However, the inclusion of other matter fields, for instance a scalar field component, will drastically alter the conclusions and possibly increase the dimensionality of the dynamical systems and/or the codimension of the unfoldings (even in cases with identical symmetries as the ones here, for instance $\mathbb{Z}_2$). Indeed, developing the versal unfolding construction for a simple modified gravity theory in, say, a Friedmann context would be an interesting problem.

Amongst further fundamental  issues emerging in the framework of gravitational bifurcation theory and deserve a comment, we shall briefly discuss  only three. Firstly, since the two appear intertwined in any bifurcation diagram,  what is the relation between the unfolding parameter (`motion' in parameter space) and the time (evolution in  phase space)? In particular, which of the two $\mu$ or $t$ is prior? Unconstrained motion in parameter space appears to be a cause for the appearance of a `time succession' in phase space (metamorphoses). Consider evolution in a stratum without equilibria, e.g., the negative axis in Fig. 1 right, or the $\chi$-stratum in Fig. 4, and the motion of a parameter point around the origin in parameter space. Passing through a bifurcation set changes the number and nature of equilibria, and forces the phase point to move on a new orbit in the `next' phase portrait. Such a smooth transition of the phase point between different phase portraits introduces a time progression into the problem associated with a global change (i.e., the appearance of one or more new equilibria). This state of affairs begs the question as to \emph{what causes motion in parameter space}? We believe that an answer to this question might be  related to the proven stability of the versal unfolding. In other words, the system collapses when the stability of the whole construction is disrupted, something which as we have shown cannot really happen in a versal unfolding.

Secondly, establishing a `general relativity theory landscape' via gravitational bifurcation theory would not just be a formal curiosity, it provides a  unifying framework that organizes the full zoo of known and yet-to-be-discovered solutions under one geometric roof (see Table \ref{string-gr-land}). While Table \ref{string-gr-land} summarizes the deep parallels between string-theory and GR-bifurcation landscapes, the rigorous derivations of each entry are beyond the scope of this review and will be presented in forthcoming work. Here we highlight just a few key features of this correspondence.  First,   a versal-unfolding landscape endows GR with a moduli space of all symmetry-reduced solutions of Einstein's equations, whose strata encode precisely how equilibrium branches appear, merge, or vanish as control parameters vary.  This construction brings systematic control of gravitational instabilities: every boundary in parameter space marks a bifurcation at which a solution gains or loses stability, ensuring that no qualitative transition is overlooked.    Second, the landscape approach bridges traditional PDE-level stability analyses and finite-dimensional ODE dynamics, extending linear results (e.g.\ around Kerr or FLRW) into a fully nonlinear classification of nearby behaviour and potentially uncovering new invariants or `charges' associated with each stratum.   Third, it forges powerful connections to algebraic-geometric and topological methods, such as discriminant varieties, resolution of singularities, and Hodge theory, providing tools to probe the global structure of solution spaces and phenomena such as crease flows on horizons. Finally, in direct analogy to string phenomenology, gravitational physicists can exploit this bifurcation landscape to guide numerical simulations and observational searches toward those regions of parameter space where novel phenomena (critical scaling laws, horizon echoes tied to global bifurcations, etc.) are most likely to emerge.

\begin{sidewaystable}[p]
  \centering
  \footnotesize
  \captionsetup{width=0.85\textwidth}
  \begin{tabular}{p{0.30\textwidth} p{0.40\textwidth} p{0.28\textwidth}}
    \hline
    \textbf{String-landscape concept} & \textbf{GR-bifurcation-landscape analogue} & \textbf{Comments} \\
    \hline
    Moduli fields & Control parameters (knobs) in a versal unfolding & Codimension = number of knobs. \\
    Vacua minima of $V(\phi)$ & Equilibrium branches $x(\lambda)$ solving $f(x,\lambda)=0$ & Stable when hyperbolic. \\
    Scalar potential $V(\phi)$ & Bifurcation function $f(x,\lambda)$ & Zeros define equilibria. \\
    Flux quantum numbers & Codimension (number of degeneracy directions) & Discrete knob choices. \\
    Stability $\nabla^2V>0$ & Hyperbolicity (invertible Jacobian) & Structural-stability condition. \\
    Tunneling & Local bifurcations (branch creation/annihilation) & Vacuum-decay analogue. \\
    Discriminant loci & Bifurcation surfaces in parameter space & Boundaries of stability regions. \\
    Metastability & Persistence of branches under small changes & Robustness measure. \\
    Anthropic selection \& dynamical selection & Basin-of-attraction volumes & Attractor preference. \\
    Calabi-Yau moduli & Versal-unfolding family & Universal deformation space. \\
   Flat (zero-frequency) directions & Center manifold & Slow-mode subspace. \\
   Transition rates computed via instantons & Normal-form analysis of bifurcations & Semi-classical tunneling \(\leftrightarrow\) normal form coefficients governing branch splitting. \\
   Vacuum-tracking / phase diagram of vacua vs.\ moduli & Bifurcation diagram & Tracks branch splitting/merging. \\
   Walls of marginal stability in moduli space & Global bifurcations & Large-scale separatrix reconnections. \\
    \hline
  \end{tabular}
  \caption{Analogy between key concepts in the string-theory landscape and their counterparts in the gravitational bifurcation landscape.}
  \label{string-gr-land}
\end{sidewaystable}

A last problem we would like to briefly discuss is related to the question of the structure and nature of the theory that contains all versal unfoldings starting from a given one. We have begun to answering this question for general relativity, but one could perhaps apply similar ideas to other areas of fundamental theoretical physics. What is the nature of the versal families that describe the unfoldings of the  Einstein, Maxwell, Schr\"{o}dinger, or Dirac equations? A way to begin to answer this question may be by looking at systems that couple gravity to such fields in a specialized context, for instance, in a isotropic universe, or a spherically symmetric star, etc.  In this sense, one arrives at a new `theory' that `pierces' through the standard ones  at  degenerate points (e.g., general relativity at the `point' (\ref{deg-sys})), and  contains all such versal final forms of the situation one started with. For example, the landscape construction of string theory may have something to do with the versal unfoldings that the original theory contains and these in turn may lead to a new classification of possible string solutions. Another example of this sort would be to start with say the Einstein-Maxwell equations in some simplified context (say for a simple Bianchi model) and unfold. What would be the nature of the resulting theory? We believe that there is a lot to learn by deploying this program for the versal unfoldings of a given theory and discovering their relations to those of a suitably chosen `neighbor'.
\addcontentsline{toc}{section}{Acknowledgments}
\section*{Acknowledgments}
The author is grateful to three anonymous referees whose suggestions and constructive comments have led to a better version of this work, and to Ignatios Antoniadis and Alexander Yefremov for useful discussions related to the work presented in this paper. This research  was funded by RUDN University,  scientific project number FSSF-2023-0003.

\end{document}